\documentclass[usegraphicx,paper]{aastex631}
\accepted{for publication in ApJ}
\shorttitle{Precise Mass Measurements of the M8-M9 binary LP~349$-$25AB}
\shortauthors{Curiel et al.}
\begin{document} 

%
% The maximum title length is 90 characters (including spaces).
%
\title{Precise Mass, Orbital motion and Stellar properties of the M-dwarf binary LP~349$-$25AB}

\correspondingauthor{Salvador Curiel}
\email{scuriel@astro.unam.mx}
\author[0000-0003-4576-0436]{Salvador Curiel$^{*}$}
\affiliation{Instituto de Astronom{\'\i}a,
Universidad Nacional Aut\'onoma de M\'exico (UNAM),
Apdo Postal 70-264,
Ciudad de M\'exico, M\'exico.
scuriel@astro.unam.mx}

%\author{et. al.}

\author[0000-0002-2863-676X]{Gisela N. Ortiz-Le\'on}
\affiliation{Instituto Nacional de Astrof\'isica, \'Optica y Electr\'onica, 
Apartado Postal 51 y 216, 
72000, Puebla, Mexico
%%Apdo Postal 70-264,
%%Ciudad de M\'exico, M\'exico.
}
%\affil{Max Planck Institut f\"ur Radioastronomie,
%Auf dem H\"ugel 69, D-53121 Bonn, Germany}}

\author[0000-0002-2564-3104]{Amy J. Mioduszewski}
\affiliation{National Radio Astronomy Observatory,
P.O. Box 0, Socorro, NM 87801, USA }

\author{Anthony B. Arenas-Martinez}
\affiliation{Facultad de Ciencias,
Universidad Nacional Aut\'onoma de M\'exico, Circuito Exterior s/n,
Ciudad Universitaria, Coyoac\'an, CDMX 04510, M\'exico.}
\affiliation{Instituto de Astronom{\'\i}a,
Universidad Nacional Aut\'onoma de M\'exico (UNAM),
Apdo Postal 70-264,
Ciudad de M\'exico, M\'exico.
scuriel@astro.unam.mx}

\begin{abstract}

\noindent

LP~349$-$25 is a well studied close stellar binary system comprised of two late M dwarf stars, both stars close to the limit between star and brown dwarf. This system was previously identified as a source of GHz radio emission.
We observed  LP~349$-$25AB in 11 epochs in 2020$-$2022, detecting both components in this nearby binary system using the Very Long Baseline Array (VLBA). 
We fit simultaneously the VLBA absolute astrometric positions together with existing relative astrometric observations derived from optical/infrared observations with a set of algorithms that use non-linear least-square, Genetic Algorithm and Markov Chain Montecarlo methods to determine the orbital parameters of the two components.
We find the masses of the primary and secondary components to be 0.08188 $\pm$ 0.00061 $M\odot$, and 0.06411 $\pm$ 0.00049 $M_\odot$, respectively, representing one of the most precise mass estimates of any UCDs to date. The primary is an UCD of 85.71$\pm$0.64 M$_{Jup}$, while the secondary has a mass consistent with being a Brown Dwarf of 67.11 $\pm$ 0.51 $M_{Jup}$. This is one of the very few direct detections of a Brown Dwarf with VLBA observations. We also find a distance to the binary system of 14.122 $\pm$ 0.057 pc.
Using Stellar Evolutionary Models, we find the model-derived stellar parameters of both stars. In particular, we obtain a model-derived age of 262 Myr for the system, which indicates that LP~349$-$25AB is composed of two pre-main-sequence stars. In addition, we find that the secondary star is significantly less evolved that the primary star.

\end{abstract}


%-------------------------------------------------------------------

\section{Introduction} \label{sec:intro}

When the UCDs are nearby (closer that 30 pc), their proximity offers many benefits, including the possibility of spatially resolving binaries with small semi-major axes and correspondingly short orbital periods. 
Precise astrometry can reveal the orbital motion of the stars in nearby binary systems, which is of particular interest, since full orbital solution of binaries is a reliable method to determine the dynamical mass of the system, as well as the masses of the individual stars \citep[][]{Baraffe15, Dupuy15}. 
Thus, close binary systems can be used to calibrate evolutionary models due to the measurable dynamical masses and coevolution of their components, which eliminates age as a confusion factor \citep[][]{Zhang20}.
Only a few close binaries with low-mass components have been studied in detail with multy-epoch Very Long Baseline Interferometry (VLBI) observations, giving high-precision mass estimates  \citep[e.g.,][]{Dupuy16, Ortiz-Leon17,Zhang20,Curiel22}. Observations of nearby UCDs also offer the possibility of finding Jupiter-like planetary companions because the astrometric signature of such planets (the reflex motion of the the star due to the gravitational pull of the companion) exceed the astrometric precision that can be achieved with the Very Long Baseline Array (VLBA) (of the order of, or even better than 100 $\mu$as). For instance, in the case of an UCD of 0.08 $M_\odot$ at a distance of 12 pc, an astrometric signal between 0.1 and 0.5 mas would be expected for a wide range of possible orbital periods $\leq$2 years and companion masses from 0.2 to 1 $M_{jup}$. If the planetary companion were more massive than Jupiter, we would expect a larger astrometric signal. If the star were closer, a larger astrometric signal would be expected.

LP~349$-$25 (LSPM~J0027+2219, 2MASS~J0027559+221932) was found to be a nearby M dwarf star by \citet[][]{Gizis00}. Later it was classified as a M8V$+$M9V close binary system (LP~349$-$25AB) with a separation of about 1.2 astronomical units (au) by \citet[][]{Forveille05}, having a rapid rotation speed of 55 $\pm$ 2 and 83 $\pm$ 3 km~s$^{-1}$, respectively  \citep[][]{Konopacky12}, and an optical rotation period of 1.86 $\pm$ 0.02 h \citep[][]{Harding13}. The mean radial velocity (RV) of the two components are $-$10.27 km~s$^{-1}$ for LP~349$-$25A and $-$6.53 km~s$^{-1}$ for LP~349$-$25B (between Dec 2006 and Jun 2009) \citep[][]{Konopacky10}. This binary has been spatially resolved with high-angular resolution optical and near infrared observations \citep[][]{Forveille05, Konopacky10, Dupuy10, Dupuy17}, showing that the system has an orbital period of about 7.7 yr, a semimayor axis of 2.1 au (0.146$''$), a nearly circular orbit (e = 0.045), a combined mass of 166 M$_{jup}$, and that it is located at a distance of 14.45 pc \citep[][]{Dupuy17}. LP~349$-$25AB (M8$+$M9) has an estimated age of $\sim$140$-$190 Myr \citep[][]{Dupuy10}, based on the estimated dynamical mass of the binary and the lack of  lithium absorption.  More recent results suggest that the binary system is a pre-main sequence binary system with a mass ratio q $\sim$ 0.94 and an estimated age of $\sim$242$-$293 Myr \citep[][]{Dupuy17}. An apparent  spin-orbit alignment of the main stellar component was found \citep[][]{Harding13}.

This binary system was detected with the Very Large Array (VLA) at 8.46 GHz by \citet[][]{Phan-Bao07}, and later at frequencies between 5 and  9 GHz \citep[][]{Osten09, McLean12}, having nearly constant flux density and very low circular polarization signal. Recently, LP~349$-$25 was detected with ALMA with a flux density of 70 $\pm$  9 $\mu$Jy, without flaring activity through the ALMA observations at 97.5 GHz \citep[][]{Hughes21}. The estimated spectral index of the binary is of $\alpha$ = $-$0.52 $\pm$ 0.05 in the frequency range between 5 and 100 GHz, which is consistent with optically thin gyrosynchrotron radiation \citep[][]{Hughes21}.

Observations of the radio emission from this binary system taken with Very Long Baseline Interferometry (VLBI) enable precise astrometry relative to extragalactic reference sources, which can be considered fixed in an inertial reference frame \citep[][]{Zhang20}.
VLBA observations can isolate the emission to individual components of the binary, and trace their absolute motion in the sky with extremely high precision. \citet[][]{Curiel22} previously employed such a method to obtain absolute astrometry, and subsequently constrain the orbits and individual masses of the M dwarf binary system GJ896AB, whose secondary component was also found to be radio emitting. Furthermore, a Jovian-like planet was found orbiting the main star of this binary system.

LP~349$-$25 presents an opportunity to investigate a system with components very close to the minimum stellar mass threshold  \citep[0.075 M$_\odot$;][]{Chabrier23}.
Here, we report 5 GHz VLBA observations of the binary LP~349$-$25AB from 11 epochs spanning 1.8 years. 
We discuss the properties of the radio emission of the two components. We then jointly fit their absolute positions together with previously published relative astrometry obtained by  \citet[][]{Forveille05},  \citet[][]{Konopacky10},  \citet[][]{Dupuy10}  and  \citet[][]{Dupuy17} to tightly constrain the absolute motion of LP~349$-$25A and B and determine their individual dynamical masses. We discuss our results in the context of earlier observations and we use evolutionary models to determine the model-derived stellar parameters of both stars.

\section{Observations and Data Reduction} \label{sec:obs}

Observations of the LP~349$-$25 binary system were obtained with the VLBA in eleven epochs, between 2020 February and 2021 December  \ref{tab_1}. Two epochs were observed at 4.85~GHz using eight 32-MHz frequency bands, in dual polarization mode, with a data rate recording of 2~Gbps. The remainder nine epochs were observed with four 128-MHz frequency bands and the 4~Gbps recording rate.

The observing sessions consisted of switching scans between the target and the phase reference calibrator, J0027+2241, spending about 1~min on the calibrator and 2~min on the target. The position of the phase reference calibrator J0027+2241 assumed during correlation was R.A.$=$00:27:15.371540 and Decl.$=$+22:41:58.06887. 
The fringe finder calibrator J0237+2848 was observed occasionally during the session. The secondary calibrators J0028+2000 and J0024+2439 were observed every $\approx$30 min and used to improve the astrometric accuracy. Additional 30-min geodetic-like blocks were observed at the beginning and end of the observing runs.

\begin{figure}[!bht]
\begin{center}
{\includegraphics[width=0.9\textwidth,angle=0]{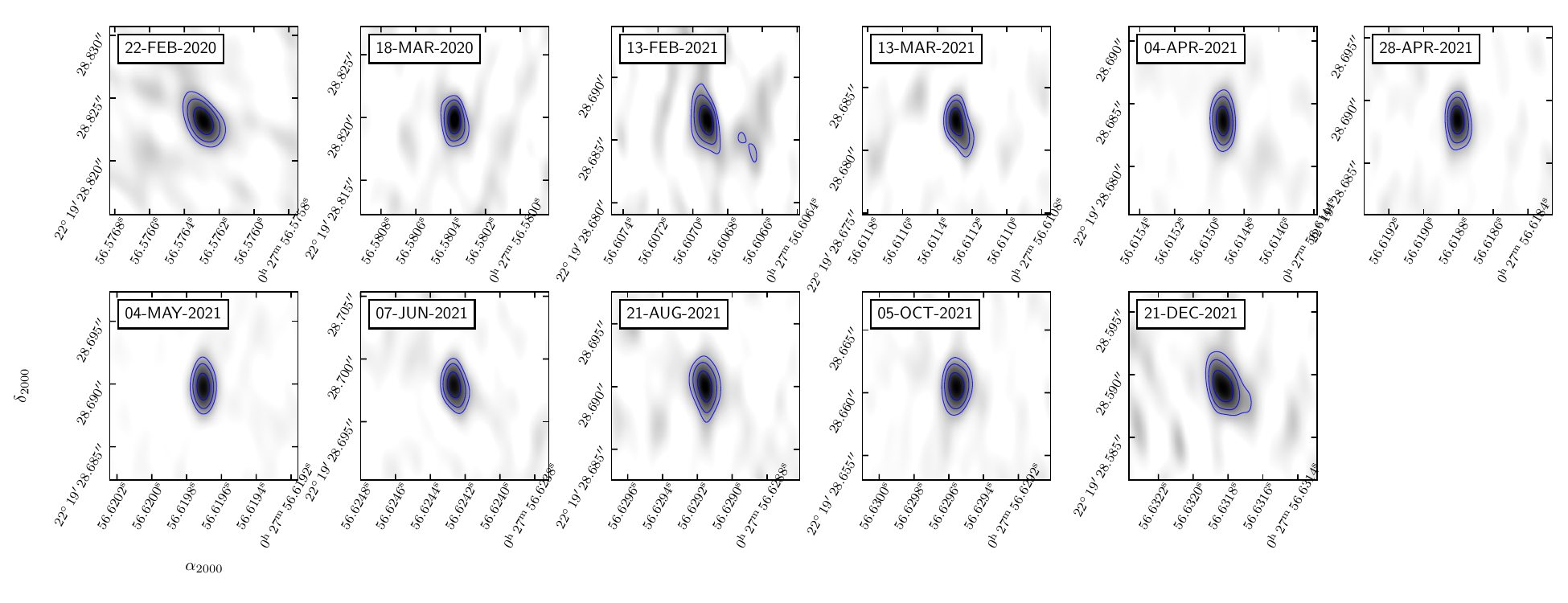}}
 \end{center}
\begin{center}
{\includegraphics[width=0.9\textwidth,angle=0]{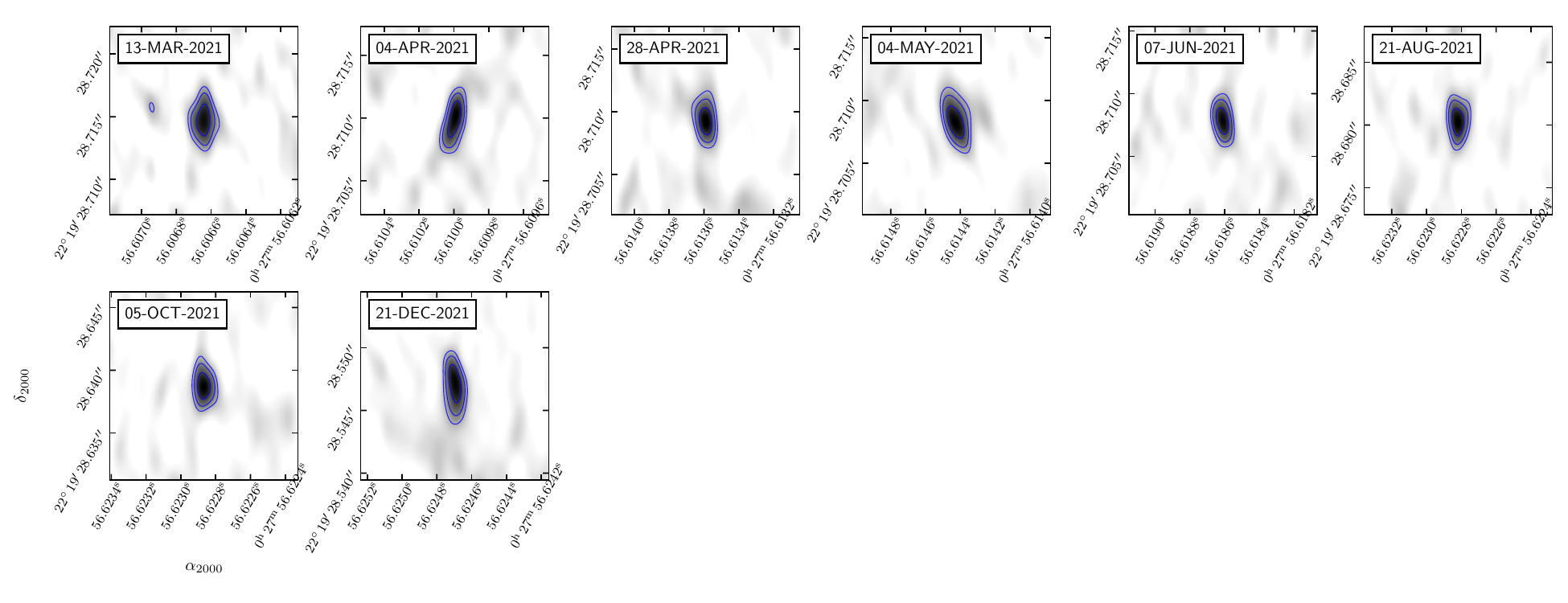}}
 \end{center}
\caption{{\bf (Upper panels:)} Intensity maps of LP--349A. The $n$th contour is at $(\sqrt{2})^2 \times S_{\rm max} \times p$ where $S_{\rm max}$ is the peak flux density, $n=0,1,2,$ and $p$ is equal to 40\%. 
{\bf (Lower panels:)} Intensity maps of LP--349B.  The contours are as above but here $S_{\rm max}$ is the peak flux density of this source. The date of observation is indicated in the legends.}
\label{fig:LP349-25-maps}
\end{figure}

We reduced the data with the Astronomical Imaging System \citep[AIPS;][]{Greisen03}, following standard procedures for phase-referencing observations \citep[][]{Torres07,Ortiz-Leon17} as described in \citet[][]{Curiel20} and \citet[][]{Curiel22}. 
First, corrections for the ionosphere dispersive delays were applied. Then, we corrected for post-correlation updates of the Earth Orientation Parameters. Corrections for the digital sampling effects of the correlator were also applied. The instrumental single-band delays caused by the VLBA electronics, as well as the bandpass shape corrections were determined from a single scan on the fringe finder calibrator and then applied to the data. Amplitude calibration was performed by using the gain curves and system temperature tables to derive the system equivalent flux density of each antenna. We then applied corrections to the phases for antenna parallactic angle effects. Multi-band delay solutions were obtained from the geodetic-like blocks, which were then applied to the data to correct for tropospheric and clock errors.  The final step consisted of removing global frequency- and time-dependent residual phase errors obtained by fringe-fitting the phase calibrator data, assuming a point source model. In order to take into account the non-point-like structure of the calibrator, this final step was repeated using a self-calibrated image of  the calibrator as a source model. Finally, the calibration tables were applied to the data and images of the target were produced using the CLEAN algorithm. We used a pixel size of 50~$\mu$as and pure natural weighting. Images of LP349-25A and LP349-25B are presented in Figure \ref{fig:LP349-25-maps}. The synthesized beam in these images are, on average, $3.9\times 1.7$~mas.

LP~349$-$25A was detected in the eleven observed epochs, while  LP~349$-$25B was detected in eight of our VLBA observations. 
To obtain the positions of the centroid in the images of LP349--25A and B, we used the task {\tt MAXFIT} within AIPS, which finds the position of the maximum peak flux density. The position error is given by the astrometric uncertainty, $\theta_{\rm res}/(2\times {\rm S}/{\rm N})$, where $\theta_{\rm res}$ is the full width at half maximum (FWHM) size of the  synthesized beam, and ${\rm S}/{\rm N}$ the signal-to-noise ratio of the source \citep{Thompson17}. In addition, we quadratically added half of the pixel size to the position error. 

In order to investigate the magnitude of systematic errors in our data, we obtain the positions of the secondary calibrator, J0028+2000, in the first eight epochs. The rms variation of the secondary calibrator position is (0.21, 0.10)~mas. The angular separation of J0028+2000 relative to the main phase calibrator is $2\rlap.{^{\rm o}}7$, while the target to main calibrator separation is $0\rlap.{^{\rm o}}4$. The main calibrator, target and secondary calibrator are located in a nearly linear arrangement.  
Since systematic errors in VLBI phase-referenced observations scale linearly with the source-calibrator separation \citep{Pradel06,Reid14}, we scale the derived position rms of J0028+2000 with the ratio of the angular separation between the target and the main calibrator to the angular separation between J0028+2000 and the main calibrator. This yields a systematic error of (0.03, 0.015)~mas, which was added in quadrature to the positions errors in each coordinate. 

Table \ref{tab_1} summarizes positions, the associated total uncertainties, and integrated flux densities of LP349$-$25A and B. The integrated flux densities were obtained by fitting the source brightness distribution with a Gaussian model. The rms values of the final maps are also given in this table. 

In total, we have the astrometric position of the primary star from 11 epochs and the astrometric position of the secondary star from 8 epochs spanning 1.8 years (see Table~\ref{tab_1}), and 20 relative astrometric positions of the secondary around the primary that have been measured in the past 18 years (see Table~\ref{tab_2}). We also include the RV of both stars (4 epochs) published by  \citet{Konopacky10}.

\section{Fitting of the Astrometric Data.}  \label{sec:procid}

We followed the same fitting procedure presented by \citet[][]{Curiel22}. 
In short, we used three astrometric fitting methods: asexual genetic algorithm AGA \citep[][]{Canto09, Curiel11, Curiel19, Curiel20}, non-linear Least-squares algorithm and MCMC \citep[][]{Curiel22}. The AGA code include iterative procedures
that search for the best fitted solution in a wide range of posible values in the multi-dimensional space of parameters. These iterative procedures help the fitting codes not be trapped in a local minimum, and to find the global minimum. In addition, we test the fitted results using different initial conditions to confirm that the best fitted solution corresponds to the global minimum solution.
This algorithm can be used  to fit absolute astrometric data (e.g., planetary systems), only relative astrometric data (e.g., binary systems), and combined (absolute plus relative) astrometric data (e.g., a planetary companion associated to a star in a binary system). To fit the astrometric data, we model the barycentric two-dimensional position of the source as a function of time ($\alpha(t)$, $\delta(t)$), accounting for the (secular) effects of proper motions ($\mu_\alpha$, $\mu_\delta$), the (periodic) effect of the parallax ($\Pi$), and the (Keplerian) gravitational perturbation induced on the host star by one or more companions, such as low-mass stars, substellar companions, or planets (mutual interactions between companions are not taken into account). 
We search for the best possible model (a.k.a, the closest fit) for a discrete set of observed data points ($\alpha(i)$, $\delta(i)$). The fitted function has several adjustable parameters, whose values are obtained by minimizing a $''$merit function$''$, which measures the agreement between the observed data and the model function. We minimize the $\chi^{2}$ function to obtain the maximum-likelihood estimate of the model parameters that are being fitted \citep[e.g.,][]{Curiel19, Curiel20}.

In addition, we also follow the non-linear Least-squares  and MCMC fitting procedures presented by  \citep[][]{Curiel22} to fit the astrometric data.
In this case, we use the open-source package {\tt lmfit} \citep[][]{Newville20}, which uses a non-linear least-squares minimization algorithm to search for the best fit of the observed data. This python package is based on the {\tt scipy.optimize} library \citep[][]{Newville20}, and includes several classes of methods for curve fitting, including Levenberg-Marquardt minimization and  emcee  \citep[][]{Foremanmackey13}. In addition, {\tt lmfit} includes methods to calculate confidence intervals for exploring minimization problems where the approximation of estimating parameter uncertainties from the covariance matrix is questionable.

The codes we use in this work include the possibility of adding RV data in the astrometric fitting. Thus, we fit simultaneously the astrometric and the RV data, which removes the ambiguity in the position angle of the ascending node ($\Omega$ and $\Omega$ + 180$^\circ$).

\section{Results} \label{sec:results}

By combining our VLBA data with published optical/IR and RV data \citep[][]{Forveille05, Konopacky10, Dupuy10,Dupuy17}, we are able to fully fit the orbital motions of the stars in the binary system LP~349$-$25AB.
The multi-epoch astrometric observations covered about 18 yrs, with an observational cadence that varies during the time observed. The observations were not spread regularly over the years, the gaps between observations ranged from weekly to monthly to 8 years (see Sec.~\ref{sec:obs}).
The time span and cadence of the observations are adequate to fit the proper motions and the parallax of this binary system, and to fit the orbital motions of the two stars around their common barycenter. 

\subsection{Genetic algorithm fit} \label{sec:caf}

\begin{figure*}
   \centering
   \plotone{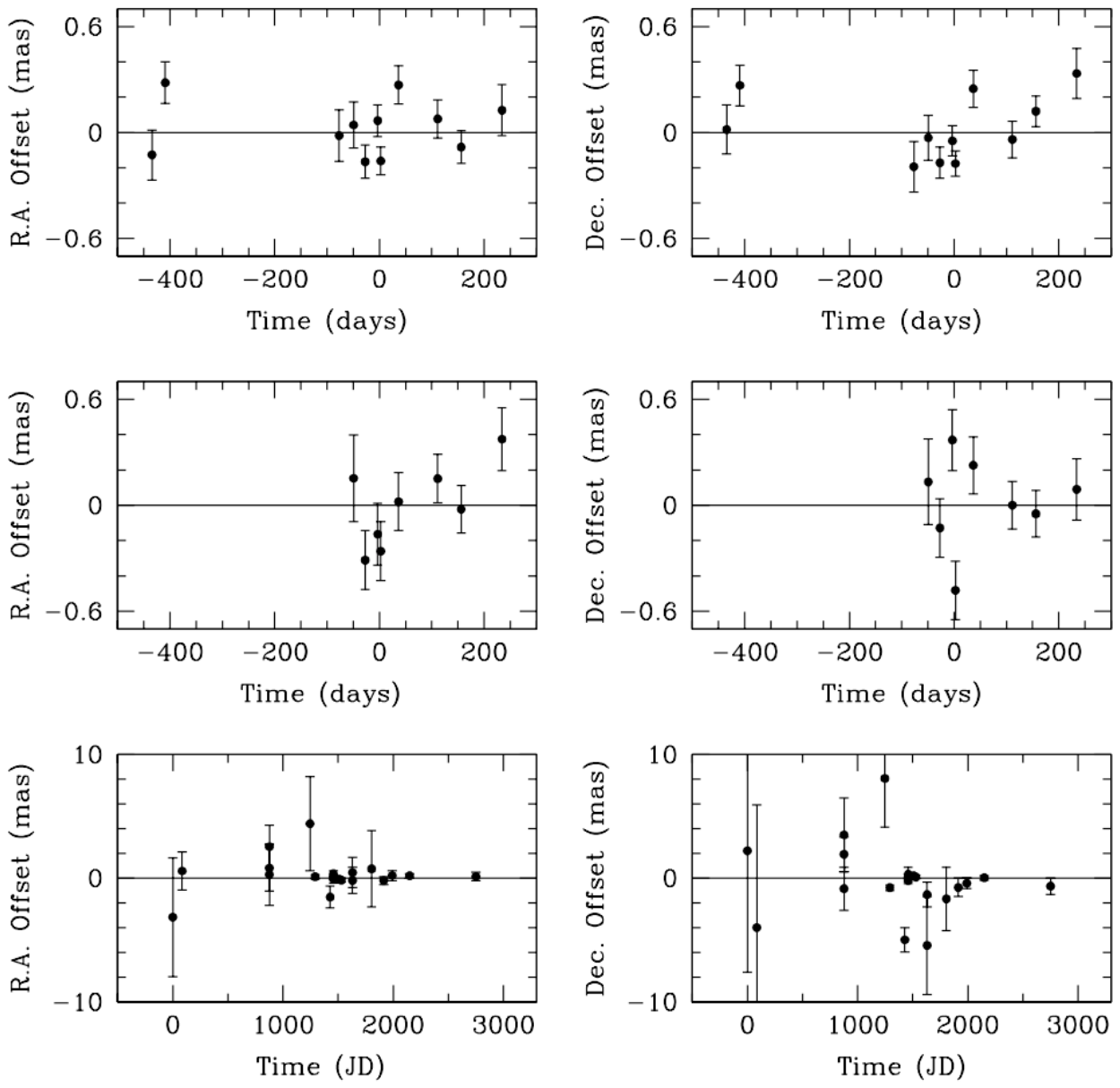}
    \caption{Residuals from the combined astrometric fit. The upper four panels show the residuals of the observed epochs of the primary star LP~349$-$25A and the secondary star LP~349$-$25B for the absolute part of the astrometric fit. The lower two panels show the residuals of the relative part of the astrometric fit.}
    \label{fig_resid}%
    \end{figure*}

The optical/infrared relative astrometry  and RV of the binary system LP~349$-$25AB   
was combined with the absolute radio astrometry of bot stars to simultaneously fit the orbital motion of both stars around their common barycenter, as well as the parallax and proper motion of the binary system.
Using the combined astrometric fit, with the asexual genetic algorithm AGA, we are able to obtain the masses of the binary and the individual stars.
The results of this combined fit are shown in Figures~\ref{fig_astromot}, \ref{fig_resid} and  \ref{fig_rv}, and summarized in column (1) of Table~\ref{tab_3}. 
We find that orbital motion of the system is nearly circular, has a semimajor axis of 2.055 au (145.52 mas) and an orbital period of 7.71 years.
The inclination angle of the orbital motion of the binary system is larger than 90$^\circ$, which indicates that the orbit is retrograde.
In addition, these results show that this binary system has a combined mass of 152.82 M$_{J}$, and that the main star is an ultra-cool dwarf of 85.71 M$_{J}$ and the secondary star is in fact a brown dwarf of 67.11 M$_{J}$.

Table~\ref{tab_3} and Figure~\ref{fig_resid} show that the residuals of the combined astrometric fit (between 0.16 and 0.26 mas) are relatively large. However, although the residuals of the fit are larger than the expected astrometric precision with the VLBA  ($<$80 $\mu$as), the residuals of the absolute astrometric fit of both stars do not show a clear temporal trend that could indicate the possible presence of close companions. 
In addition, the reduced $\chi^{2}$ value of the fit is larger than one, which may suggests that the formal errors of the observations could be underestimated, or that the fitting model is incomplete (for instance, the stars may have low mass companions). In Table~\ref{tab_3}, we have scaled up the errors of the fitted parameters by the square root of the reduced $\chi^{2}$ value. The possible presence of close companions will be discussed elsewhere.

\subsection{Non-linear least squares and MCMC fits} \label{sec:mcmc}

\begin{figure*}
   \centering
    \plotone{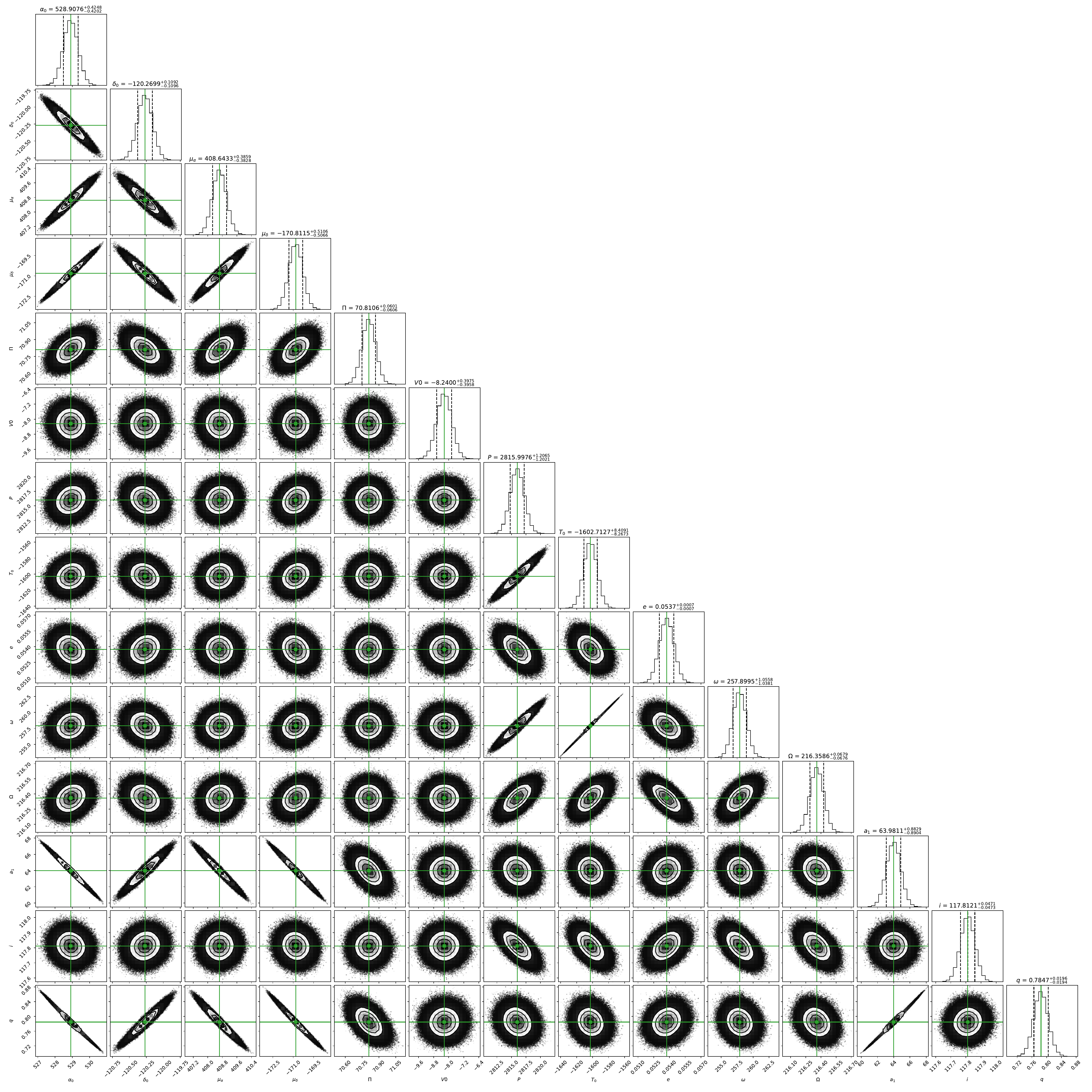}
    \caption{Posterior distributions of the fitted parameters. Combined astrometric fit of the UCD binary system LP~349$-$25AB using {\tt lmfit}. This figure shows the correlations between the fitted parameters from the MCMC analysis using the corner code. The 2D posterior probability histogram of each fitted parameter is shown on top of each column. The green lines indicate the mean value of each fitted parameter, and the two dotted vertical lines represent the $1\sigma$ range of the distribution.}
    \label{fig_emcee}%
    \end{figure*}

We used the open-source package $lmfit$ \citep[][]{Newville20}, which includes several minimization algorithms to search for the best fit of observational data. In particular, we used the default Levenberg-Marquardt minimization algorithm, which uses a non-linear least-squares minimization method to fit the data. This gives an initial fit solution to the combined astrometric bidimensional and RV data.
{\tt lmfit} also includes a wrapper for the Markov Chain Monte Carlo (MCMC) package {\tt emcee} \citep[][]{Foremanmackey13}. When fitting the combined astrometric and RV data,
 we weighted the data by the positional errors of both coordinates ($\alpha$ and $\delta$) and the RV errors. We used 250 walkers and run the MCMC for 30000 steps with a 1000 step burn-in, at which point the chain length is over 50 times the integrated autocorrelation time. 
 The fitted solutions are listed in columns (2) and (3) in Table~\ref{tab_3}, and Figure~\ref{fig_emcee} shows the correlation between the fitted parameters. The fitted solutions are very similar to those obtained from the combined astrometric fit (see column (1) in Table~\ref{tab_3}). The $\chi^{2}_{red}$ and the residuals of the fit are also very similar to those obtained from the combined astrometric fit. The errors of the fitted parameters included in columns (2) and (3)  of Table~\ref{tab_3} were scaled up  by the square root of the reduced $\chi^{2}$ value.

\section{Discussion}  \label{sec:discusion}

\subsection{Comparison with published Proper Motions and Distance of the Binary System} \label{sec:pmot}

\begin{figure*}
\centering
   \plotone{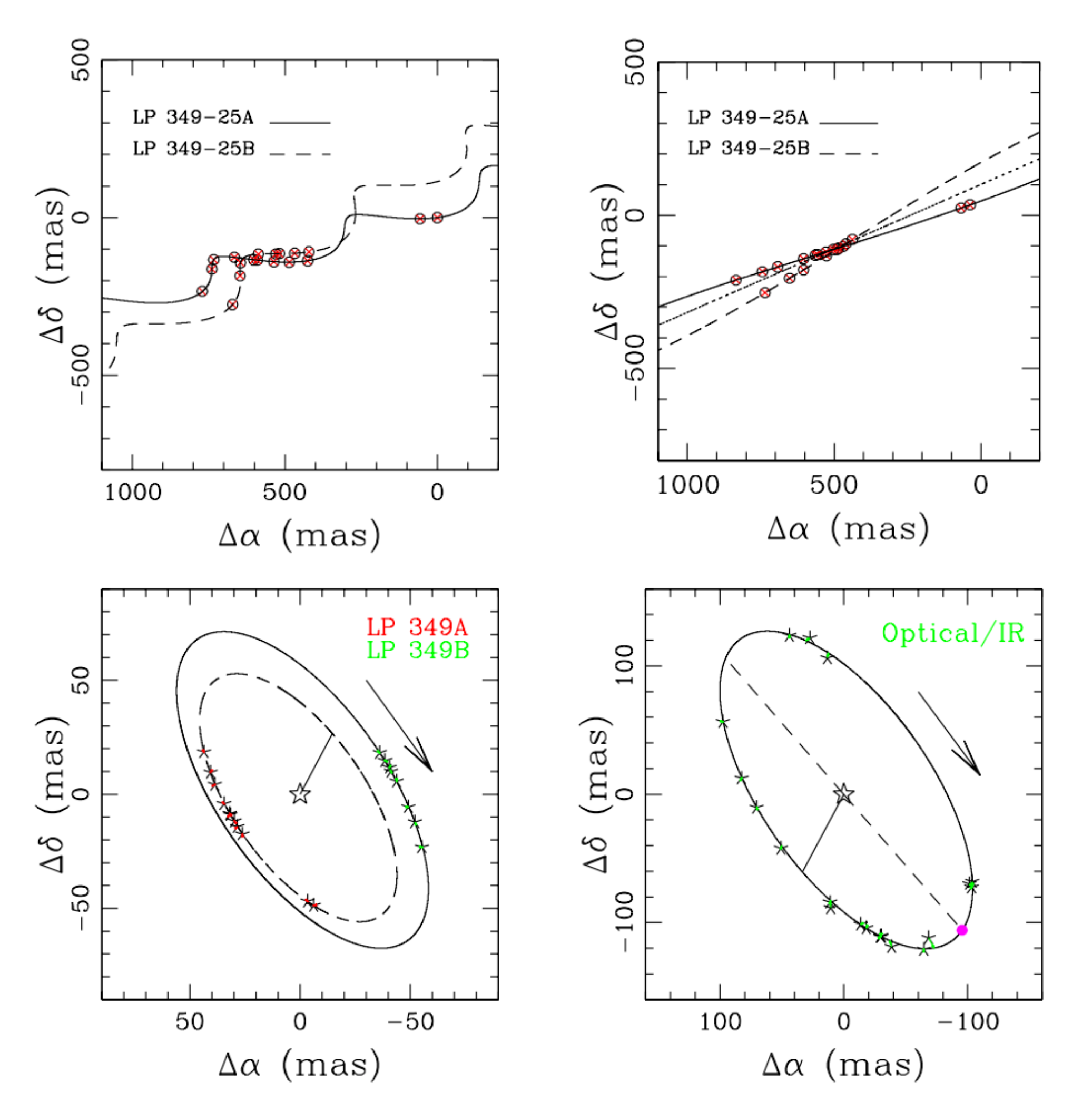}
    \caption{{\bf (Upper-Left panel:)} Absolute astrometric fits of the astrometric positions of UCD binary system LP~349$-$25AB, obtained with the VLBA. The fits includes only proper motions, parallax and the orbital motion of both stars around the barycenter of the binary system. {\bf (Upper-Right panel:)} Same as upper-left panel but removing the contribution of the parallax. The dotted line shows the trajectory (from NW to SE) of the barycenter of the binary system. \\
   Combined astrometric fit of the M Dwarf binary system LP~349$-$25AB. {\bf (Lowe-Left panel:)} Orbital motion of both UCD stars around the center of mass of the binary system. The inner and outer elipses show the orbital motion of the primary star LP~349$-$25A and the secondary star LP~349$-$25B, respectively. The VLBA observations cover approximately 20$\%$ of the orbit. The arrow shows the direction of the orbital motion. The straight line indicates the position of the periastron of the primary around center of mass.
    {\bf (Lower-Right panel:)} Relative orbital motion of the secondary star LP~349$-$25B around the primary star LP~349$-$25A. The optical/infrared observed epochs are shown in green. The arrow shows the direction of the orbital motion. The temporal distribution of the observations covers the full relative orbit of the binary system. The straight line indicates the position of the periastron in the relative orbit. The dotted line shows the location of the ascending (filled magenta circle) and descending nodes.}   
    \label{fig_astromot}%
    \end{figure*}

The estimation of the proper motions and the distance of a binary system is complex due to the orbital motions of each component around their common barycenter, specially when both stars have different masses ($m_{B}/m_{A} <$ 1) and the time span of the observations cover only a fraction of the orbital period of the binary system. The best way to separate the orbital motion and the proper motion of the system is by simultaneously fit the proper motions, the parallax, and the orbital motion of the binary system. The combined astrometric fit that we carried out (see Sec.~\ref{sec:caf}) includes all these components. 
Although the relative astrometric observations cover the full orbit of the binary system and the absolute astrometric observations cover only a fraction of the binary orbit (see Figure~\ref{fig_astromot}), we obtain an excellent solution for the combined astrometric fit of the binary system.
The combined astrometric solution (see column (1) of Table~\ref{tab_3}) shows that the orbital motions of the two stars around the center of mass of the system are well constrained (see Figure~\ref{fig_astromot}). 
Thus, this combined fit gives an excellent estimate of the proper motion and the parallax of the barycenter of the binary system, and the orbital motion of both stars around the barycenter (see Table~\ref{tab_3}). 

GAIA observations do not resolve the binary system. 
The DR3 catalog of GAIA gives the estimated proper motions and parallax of this binary system $\mu_{\alpha}$ = 392.72 $\pm$ 0.47 mas yr$^{-1}$,  $\mu_{\delta}$ = $-$186.59 $\pm$ 0.40 mas yr$^{-1}$, and $\Pi$ = 70.78 $\pm$ 0.43 mas. 
The solution that we obtain is $\mu_{\alpha}$ = 408.68 $\pm$ 0.40 mas yr$^{-1}$, $\mu_{\delta}$ = $-$170.77 $\pm$ 0.39 mas yr$^{-1}$, and $\Pi$ = 70.81 $\pm$ 0.29 mas. 
Comparing our combined astrometric solution with that obtained by GAIA we obtain that: 
$\Delta\mu_{\alpha}$ = 15.96 $\pm$ 0.62 mas yr$^{-1}$,
$\Delta\mu_{\delta}$  = 15.82 $\pm$ 0.56 mas yr$^{-1}$,
$\Delta\Pi$  = 0.03 $\pm$ 0.52 mas.
The combined astrometric fit solution is somewhat different from that obtained by GAIA.  In particular, the proper motions of the system obtained with the combined fit are significantly better than those obtained by GAIA. 
The different solutions are probably due to the fact that our fit uses astrometric data that cover a couple of years of the absolute astrometric positions of both stars and several orbits of the binary system, while the GAIA fit uses astrometric data obtained during the first 2.8 years of GAIA observations, which only covers about 36\% of the orbital period of the binary system. In addition, GAIA observations do not resolve the binary system and thus the photo-center of the GAIA observations is located somewhere between the position of both stars, and probably does not coincide with the barycenter of the binary system, which adds an extra movement to the photo-center due to the orbital motion of the binary system. 
We notice that the estimated parallax that we obtain is consistent with that obtained by GAIA within the estimated errors. However, the parallax that we obtain, with an estimated error of only 0.41\%, is an improvement to that obtained by GAIA, with an estimated error of 0.61\%.

\citet[][]{Dupuy17} fitted the proper motion and parallax of the system together with the orbital motion of the secondary star LP~349$-$25B around the primary star LP~349$-$25A. They estimated proper motions of the system $\mu_{\alpha}$ = 407.9 $\pm$ 1.7 mas yr$^{-1}$,  $\mu_{\delta}$ = $-$170.4 $\pm$ 1.3 mas yr$^{-1}$, which are consistent, within the errors, with the values that we find here. However, the proper motions that we obtain are more precise, and thus, our astrometric fit provides a significant improved solution to the proper motions of the binary system.  In addition, \citet[][]{Dupuy17} obtain a parallax of $\Pi$ = 69.2 $\pm$ 0.9 mas, which differs from our solution by $\Delta\Pi$  = 1.61 $\pm$ 0.95 mas. This difference is probably due to: (a) \citet[][]{Dupuy17} use unresolved astrometry of the binary system to determine the parallax and proper motions, as well as to constrain the photocenter motion due to the binary orbit, while our observations resolve both stars, and we obtain multi-epoch positions of each star, and (b) their integrated-light astrometry was obtained between the months of July and December \citep[][see their Table 4]{Dupuy17}, less than half the orbit of earth around the sun, which may affect the fit of the parallax ellipse. Thus, the parallax (and the distance) that we obtain here is a significant improvement to those previously obtained.

\subsection{Comparison with published Orbits and Mass Ratios} \label{sec:comorb}

LP~349$-$25 was found to be a nearby low-mass binary system by \citep[][]{Gizis00}. Since then, this system has been subject of several astrometric studies. In particular, in the past two decades precise optical/IR observations have resolved this binary, providing the angular separation and position angle of both stars. In the past decade it has been posible to obtain astrometric fits of the orbit of this binary system. Here, we compare the precise astrometric fit that we obtain with those obtained by \citet[][]{Konopacky10}, \citet[][]{Dupuy10} and \citet[][]{Dupuy17} (see  Table~\ref{tab_4}). 

The fitted orbital parameters that we obtain are in general similar to those obtained previously. However there are some important differences in the different fits. In particular, the position angle of the ascending node ($\Omega$) and the longitude of the periastron ($\omega$) obtained here are similar to those obtained by \citet[][]{Konopacky10}, but differ from those obtained by \citet[][]{Dupuy10} and \citet[][]{Dupuy17} by nearly 180$^\circ$ (see Table~\ref{tab_4}). 
This difference is probably because \citet[][]{Dupuy10} and \citet[][]{Dupuy17} did not take into account the radial velocity of both stars in their orbital fit.
Other important differences are the total mass of the binary system and the mass ratio of the stars. 
We find a total mass for this binary system of 0.1460 $\pm$ 0.0007 M$_{\sun}$, which disagrees with the total mass of  0.121 $\pm$ 0.009 M$_{\sun}$ obtained by \citet[][]{Konopacky10}, 0.120$^{+0.008}_{-0.007}$  M$_{\sun}$ obtained by \citet[][]{Dupuy10}, and 0.158$^{+0.006}_{-0.007}$ M$_{\sun}$ obtained by \citet[][]{Dupuy17}. This difference is mainly due to the parallax used to estimate the total mass. \citet[][]{Konopacky10} and \citet[][]{Dupuy10} used a fixed parallax of 75.8 mas, while \citet[][]{Dupuy17} obtained a parallax of 69.2 mas, which is closer to the parallax we obtain (70.810 mas). 

The mass ratio that we obtain is significantly different from those previously obtained (see  Table~\ref{tab_4}).
With the astrometric fit we obtain the dynamical mass of both stars that correspond to a mass ratio $q$ = 0.785 $\pm$ 0.029. 
\citet[][]{Konopacky10} found a model-derived mass ratio $q$ = 2, which implies that the secondary star LP~349$-$25B is more massive than the primary star LP~349$-$25A. 
Using the total mass of the binary system and the bolometric luminosity of each star, \citet[][]{Dupuy10} obtained a model-derived mass ratio $q$ = 0.87.
More recently, \citet[][]{Dupuy17} reported a mass ratio of $q$ $=$ 0.941$_{-0.030}^{+0.029}$ obtained from their astrometric fit, which is quite different to the mass ratio that we obtain.
\citet[][]{Dupuy17} also reported a model-derived mass ratio $q$ = 0.88, which was obtained by using the total mass of the system and the bolometric luminosities of each star. 
The model-derived mass ratio obtained by \citet[][]{Dupuy10} and \citet[][]{Dupuy17} are close to the dynamical mass ratio that we obtain from our astrometric fit. 
We notice that we also obtain a model-derived mass ratio of $q$ = 0.88 when using the total dynamical mass of the system that we obtain together with the bolometric luminosity of each star obtained by \citet[][]{Dupuy17} (see discussion bellow: Sec.~\ref{sec:evolmod}).

\subsection{Expected Radial Velocities} \label{sec:radvel}

\begin{figure}[!bht]
\begin{center}
{\includegraphics[width=0.6\textwidth,angle=0]{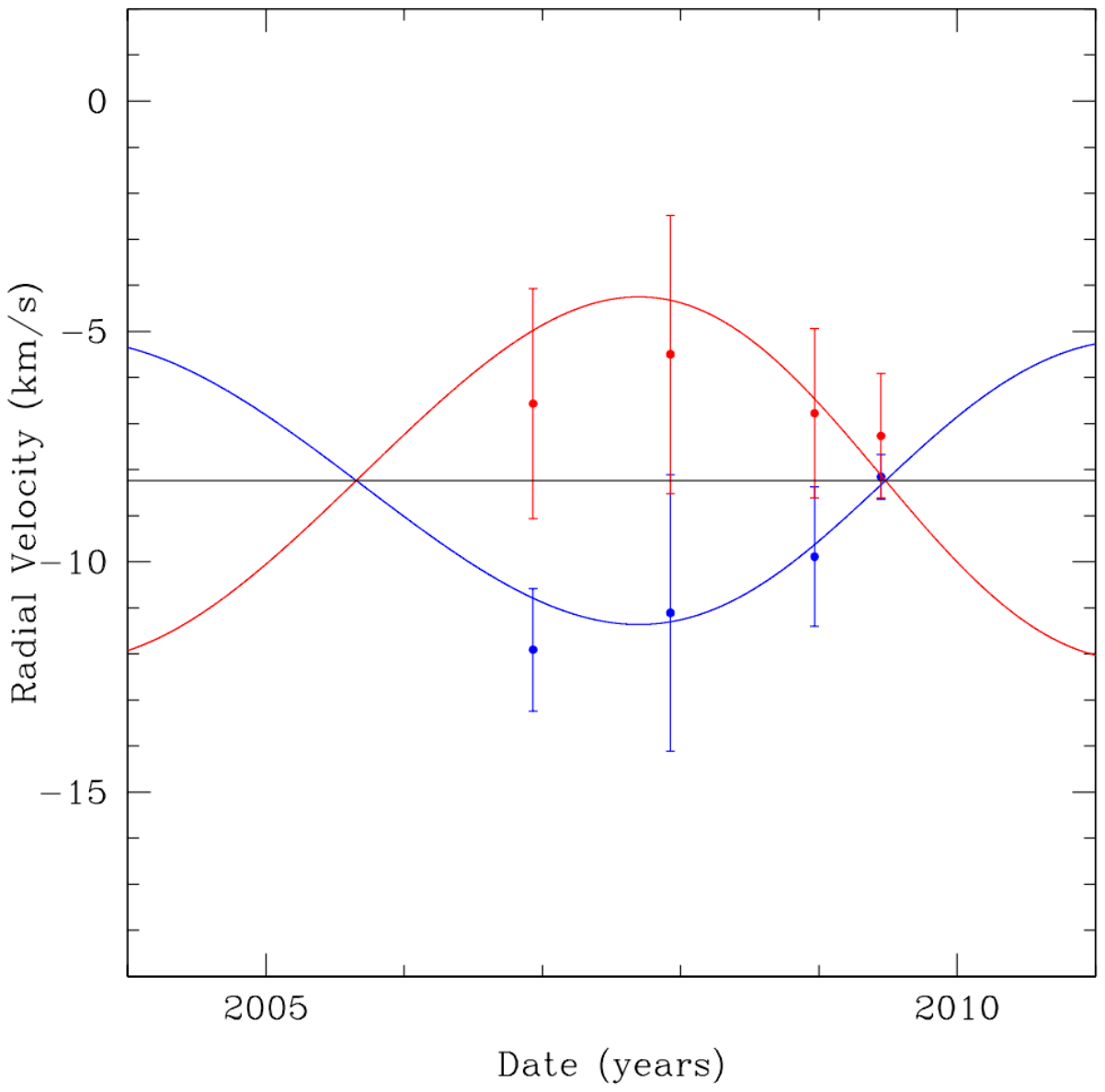}}
 \end{center}
 \caption{Radial velocity curves of LP~349$-$25A (blue) and  LP~349$-$25B (red). The solid lines correspond to the model radial velocity of both stars as function of time, obtained using the solution of the combined astrometric fit and assuming that the systemic velocity of the binary system is $-$8.24 km~s$^{-1}$. The dots show the radial velocity of both stars obtained at 4 epoch \citep[][]{Konopacky10}.
 }
 \label{fig_rv}%
\end{figure}

The solution of the combined astrometric fit can be used to estimate an expected induced maximum radial velocity (RV) of each star due to the gravitational pull of its companion as follows \citep[e.g.,][]{Canto09,Curiel20}:

\begin{equation}
K =\left(\frac{2 \pi G}{P}\right)^{1/3} \frac{m_{B} sin(i)} {(m_{B} + m_{A})^{2/3}} \frac{1} {\sqrt{1 - e^{2}}},
\end{equation}

\noindent
where G is the gravitational constant, and $P$, $m_{A}$, $m_{B}$, and $e$ are the estimated orbital period, primary and seconday masses, and the eccentricity of the orbit of the companion. 
Using the combined astrometric solution (see column (1) of Table~\ref{tab_3}), the maximum RV of LP~349$-$25A induced by the stellar companion LP~349$-$25B is $K_{A}$ $\sim$ 3.088 km s$^{-1}$, and the maximum RV of LP~349$-$25B induced by  LP~349$-$25A is $K_{B}$ $\sim$ 3.944 km s$^{-1}$.
We can also obtain the RV curve of both stars using the solution of the combined astrometric fit \citep[][]{Green93}:

\begin{eqnarray}
V_{A} = V_{0} + K_{A} \left[cos(\nu + \omega) + e ~cos(\omega)\right],  \nonumber \\
V_{B} = V_{0} -  K_{B} \left[cos(\nu + \omega) + e ~cos(\omega)\right],
\end{eqnarray}

\noindent
where V$_{0}$ is the systemic velocity of the binary system, K$_{A}$ and  K$_{B}$ are the radial velocity semi-amplitudes of both stars along the line of sight,  $\nu$ is true anomaly, $\omega$ is the longitude of the periastron of the primary star, and $e$ is the eccentricity of the orbit. 
Figure~\ref{fig_rv} shows the observed radial velocities \citep[][]{Konopacky10}  on top of the radial velocity curves of both stars. The radial velocity of the binary system, obtained by the combined astrometric fit, is V$_{0}$ = $-$8.24 $\pm$ 2.47 km~s$^{-1}$. This figure shows that the radial velocity curves follow reasonably well the observed RVs.
 
The maximum radial velocity of both stars occurred in 2007.6943 (August 2007), 2015.404 (June 2015) and 2023.114 (February 2023), when the secondary star LP~349$-$25B passed through the   ascending node of its orbits around the barycenter of the binary system (see Figure~\ref{fig_rv}).

\subsection{Flux variability of the source} \label{sec:fluxvar}

LP~349$-$25AB  is an unusual ultra-cool dwarf binary system. Low resolution, multi-epoch, centimeter radio observations of  LP~349$-$25AB have shown that the radio emission of this system is quiescent and with a constant spectral index, with no evidence of flaring or variability \citep[][]{Phan-Bao07,Osten09,McLean12}. This system was also detected at millimeter wavelengths (92 GHz) with ALMA, showing that the millimeter emission of this system is also quiescent over time spans of 2 hrs, with no evidence of flaring or variability. It was also found that the system has a spectral index $\alpha$ = $-$0.52 between 5 GHz and 92 GHz, consistent with optically thin gyrosynchrotron radiation \citep[][]{Hughes21}.

\begin{figure*}
   \centering
    \plotone{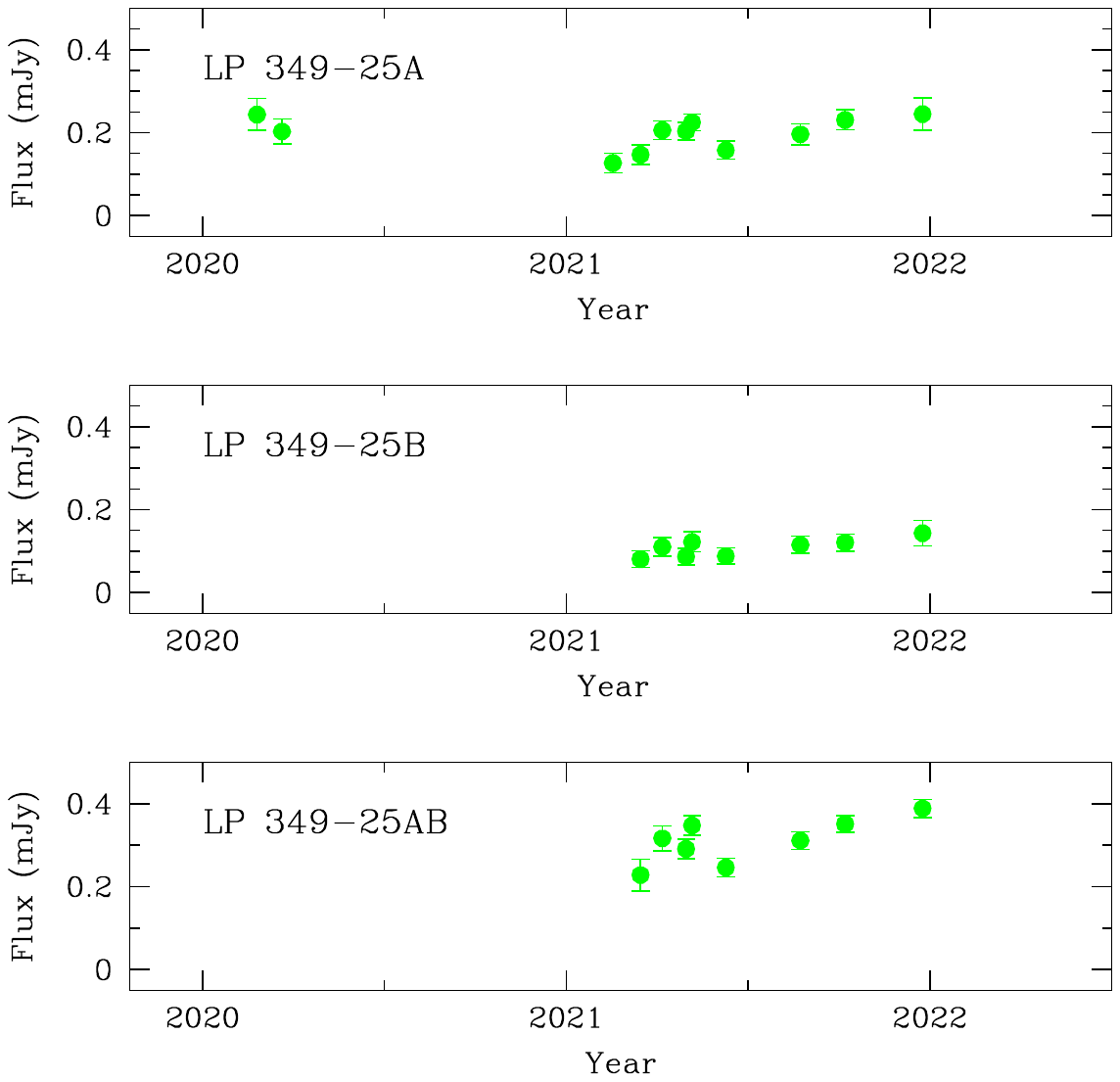}
    \caption{The upper panel shows the integrated flux density of the primary star LP~349$-$25A. The middle panel shows the same but for the secondary star LP~349$-$25B. The lower panel shows the integrated flux density of the UCD binary system LP~349$-$25AB.}
    \label{fig_flux}%
    \end{figure*}

Our VLBA observations of this system show that both stars have nearly constant  flux densities at time spans of a few hours and several months (see Figure~\ref{fig_flux}). 
The mean flux density of the primary and secondary stars are 0.20 $\pm$ 0.04 and 0.11 $\pm$ 0.02 mJy, respectively.
In addition, the mean total flux density of the binary system ($\sim$ 0.29 $\pm$ 0.04 mJy) is consistent with the estimated flux density from low angular resolution observations of this system obtained at the same frequency ($\sim$ 0.33 $\pm$ 0.04 mJy; \citet[][]{Osten09}). 
This suggests that the binary system, although both stars having a rapid rotation speed (55 $\pm$ 2 and 83 $\pm$ 3 km~s$^{-1}$; \citet[][]{Konopacky12}), very short optical rotation period (1.86 $\pm$ 0.02 h; \citet[][]{Harding13}), and being ultra-cool dwarfs, does not show outbursts or strong time variability at short (hours) and large (years) time scales. 
This also suggests that the radio emission is compact and that we are not resolving out the flux emission with out VLBA observations.
However, a close look to the temporal distribution of the flux density of each star shows that LP~349$-$25B has a nearly constant flux density as function of time, while LP~349$-$25A seems to have a small variation in time, reflected in the larger standard deviation in the flux density of this star. Figure~\ref{fig_flux} shows a small temporal fluctuation as function of time, having a nearly sinusoidal patters with a minimum of $\sim$0.1 mJy and a maximum of $\sim$0.25 mJy.
A similar temporal pattern can be observed in the integrated flux density of the binary system. 
However, it is not clear if this flux density variation is periodical. 
Further observations will be required to find if the flux variability has a defined temporal period.

\subsection{Comparison with Evolutionary Models} \label{sec:evolmod}

Direct measurements of the dynamical mass of the individual stars in a binary system enables unique tests of theoretical models of very low mass stars and brown dwarfs. 
Given the precise parallax and  individual masses of both stars in the binary system LP~349$-$25AB that we obtain with the combined astrometric fit, we can infer other physical properties of the stars from evolutionary models. In order to constrain the evolutionary models, we use the individual luminosities of both stars obtained by \citet[][]{Dupuy17}, as well as the dynamical mass of the individual stars that we obtain. 
We consider three families of evolutionary models here: \citet[][hereinafter BHAC15]{Baraffe15}, \citet[][hereinafter CLES-solar]{Fernandes19}, and \citet[][hereinafter ATMO20-CEQ]{Phillips20}. The  BHAC15 are the most recent grids from the Lyon group with a grid sample of masses adequate for brown dwarfs and low mass star (0.01 $M_{\odot}$ to 1.4 $M_{\odot}$). The CLES-solar are the standard models for solar abundance, with a grid sample of masses adequate for early brown dwarfs and very low mass star (0.055 $M_{\odot}$ to 0.13 $M_{\odot}$). We use the models with equilibrium chemistry of ATMO20-CEQ with a grid sample of masses adequate for substellar objects (0.0005 $M_{\odot}$ to 0.075 $M_{\odot}$).

Here, we use three methods to find the model-derived stellar parameters:

\begin{itemize}

{\item {\bf Method 1: Combined Mass and Individual Luminosities.}} 
In this first method, we  use three different constrains during the fit of the BHAC15 evolutionary  models: 
(a) we assume that the sum of the model-masses of the components is equal to the dynamical mass of the binary system that we obtain here (see Table~\ref{tab_3}), (b) the model-derived luminosity of each star is equal to that obtained by \citet[][]{Dupuy17}, and  (c), we assume that both stars are coeval. We draw random values of mass for the primary star and derive the mass of the secondary star using the mass of the binary system (m$_{B}$ = m $-$ m$_{A}$) for each step in our Monte Carlo in-house code. Then, we bilinearly interpolate each resultant pair of (age, L$_{bol}$) for each star. With the restrictions imposed, the code converges rapidly to a single solution, regardless of the initial age used in the fit. Once we find the model-derived cooling age of the system, we repeat the process to estimate the other stellar parameters by bilinearly interpolating each resultant pair of parameters, such as (age, T$_{eff}$). This procedure provides the best fit of the model-derived stellar parameters. To obtain an estimation of the errors, we use the estimated error of the individual luminosity and the estimated error of the total mases of the binary system. 

The resulting BHAC15 model-derived values of  stellar mass, cooling age, T$_{eff}$, radius, log $g$, and fraction of lithium remaining (Li/Li$_{init}$) are summarized in Table~\ref{tab_5}. We find that model-derived cooling age of the binary system 230$\pm$16 Myr. Other model-derived parameters are obtained for each star in the binary system: M$_{A,B}$ = 0.0777, 0.0683 M$_{\odot}$ ($q = 0.88$), T$_{eff}$ = 2699 $\pm$ 17, 2574 $\pm$ 19 K and Li/Li$_{init}$$<< 1\%$, $<< 1\%$. We find that both stars are predicted to be fully depleted in Lithium. This result is consistent with the absence of Lithium absorption in this system \citep[][]{Reiners09}. The small formal uncertainties in our model-derived parameters reflect the precision of the measured masses and luminosities projected onto the model grids; we do not attempt to include any systematic errors that could be associated with the models.

{\item {\bf Method 2: Individual Masses and combined Luminosity.}} In this second method, we use three constrains during the fit of the BHAC15 and CLES-solar evolutionary models: (a) the model-mass of each star is equal to the dynamical mass of the star (see Table~\ref{tab_3}), (b) we assume that the sum of the model-derived luminosities of the components is equal to the total luminosity of the binary system obtained by \citet[][]{Dupuy17}, and (c) we assume that both stars are coeval. We first bilinearly interpolate the mass of each star in the evolutionary models to obtain the corresponding grid of models for a star with a mass equal to the dynamical mass of each star. We draw random values of age of the system for each step in our Monte Carlo in-house code. Then, we bilinearly interpolate each resultant pair of (age, L$_{bol}$) for each star. With the restrictions imposed, the code converges rapidly to a single solution, regardless of the initial age used in the fit. Once we find the model-derived cooling age of the system, we repeat the process to estimate the other parameters by bilinearly interpolating each resultant pair of parameters, such as (age, T$_{eff}$). This procedure provides the best fit of the model-derived stellar parameters. To obtain an estimation of the errors, we use the estimated error of the total luminosity and the estimated error of the mases of both stars. 

The resulting BHAC15 and CLES-solar model-derived values of cooling age,  L$_{bol}$, T$_{eff}$, radius, log $g$, and fraction of lithium remaining (Li/Li$_{init}$) are summarized in Table~\ref{tab_5}. We find that model-derived cooling age of the binary system from BHAC15 is 232$\pm$15 Myr and from CLES-solar is 224$\pm$16 Myr. These cooling ages differ by only $\sim$8 Myr, but they are consistent within the estimated uncertainties. 
Thus, BHAC15 and CLES-solar evolutionary models give the same cooling ages for both stars.
The model-derived bolometric luminosity of each star is similar:  log($L_{bol}/L_{\odot}$) = $-$3.050 dex for the main star and $-$3.286 dex for the secondary star.
The model-derived effective temperature of the main and secondary stars differ by 12$^{\circ}$ and 16$^{\circ}$, respectively. These differences are also consistent with the estimated uncertainties.
The model-derived stellar radius and gravity of each star are also similar and consistent with the estimated uncertainties. 
Both evolutionary models indicate that the primary star is fully depleted in Lithium. However, they suggest that in the secondary star remains between 0.3$\%$ and 0.5$\%$ of its initial Lithium. This result is consistent with the absence of Lithium absorption in this system \citep[][]{Reiners09}.

{\item {\bf Method 3: Individual Masses and Individual Luminosities.}} In this third method, we  use two constrains during the fit of the BHAC15, CLES-solar and ATMO20-CEQ evolutionary models: 
(a) the individual model-mass of each star is equal to the dynamical mass of the star (see Table~\ref{tab_3}), and (b) the individual model-luminosity of each star is equal to the luminosity of the star obtained by \citet[][]{Dupuy17}. We first bilinearly interpolate the mass of each star in the evolutionary models to obtain the corresponding grid of models for a star with a mass equal to the dynamical mass of each star. We then bilinearly interpolate the pair (L$_{bol}$, age) for both stars. With the restrictions imposed, the code gives a single solution. Once we find the model-derived cooling age of the system, we repeat the process to estimate the other stellar parameters by bilinearly interpolating each resultant pair of parameters, such as (L$_{bol}$, T$_{eff}$). This procedure provides precise  model-derived stellar parameters. To obtain an estimation of the errors, we use the estimated error of the individual luminosity and the individual masses of each star. 
The resulting model-derived values of cooling age, T$_{eff}$, radius, log $g$, and fraction of lithium remaining Li/Li$_{init}$ are summarized in Tables~\ref{tab_5} and \ref{tab_6}. 

For the primary star, we find that BHAC15 and CLES-solar give model-derived cooling ages of 262$\pm$21 Myr, and 255$\pm$19 Myr, respectively. These cooling ages differ by only $\sim$ 7 Myr, which is consistent within the estimated errors.
In the case of the other stellar parameters, both evolutionary models give basically the same model-derived effective temperature, stellar radius, and gravity, within the estimated errors. In addition, both evolutionary models indicate that the primary star is fully depleted in Lithium.

For the secondary star,  BHAC15 and ATMO20-CEQ evolutionary models give the same cooling age of 199 $\pm$ 12 Myr, while the CLES-solar evolutionary models give a cooling age of 191 $\pm$ 12 Myr. The difference in the cooling age is of about 8 Myr, which is within the estimated uncertainties.
ATMO20-CEQ evolutionary models give a significantly higher effective temperature (2585 $\pm$ 20 K) than BHAC15 and CLES-solar evolutionary models (2553 $\pm$18 K and 2536 $\pm$ 19 K, respectively).
The three evolutionary models give similar stellar radius and gravity.
BHAC15 evolutionary models indicate that the secondary star still has about 0.7$\%$ of its initial Lithium, while CLES-solar evolutionary models suggest that this young brown dwarf still has about 2.5$\%$ of its initial Lithium. 
These depleted fractions are consistent with the no detection of Lithium absorption in the integrated light of this binary system \citep[][]{Reiners09}.

\end{itemize}

The model-derived parameters of LP~349$-$25A and  LP~349$-$25B obtained with the three  methods significantly differ from each other. The main difference reside in the model-derived cooling age of the individual stars, and the masses and luminosities of the individual stars compared to those obtained from the observations. The first method estimates the mass of the individual stars using as restriction the total mass of the system, the individual luminosities and that both stars formed simultaneously. The model-derived mass of the primary star (0.0777 $M_{\odot}$) and the secondary star (0.0682 $M_{\odot}$) are significantly lower and higher, respectively, than the  dynamical masses that we obtain for both stars (0.0819 and 0.0641 $M_{\odot}$).
Thus, the model-derived mass ratio ($q$ = 0.88) is significantly larger that the one we obtain ($q$ = 0.783). The model-derived luminosity of LP~349$-$25A and  LP~349$-$25B (log(L) = $-$3.050 and $-$3.286 L$_{\odot}$) are significantly higher and lower, respectively,  than those obtained by \citet[][]{Dupuy17} ($-$3.075 and $-$3.198 L$_{\odot}$).

The first two methods, where we have assumed that both stars were coeval, provide, as expected, basically the same age for the binary system  (232 and 224 Myr). However, with the third method, were the dynamical masses and luminosities of the individual stars were used as constrains, and we have not assumed that the two stars were coeval, we obtain a different age for LP~349$-$25A and  LP~349$-$25B (262 and 199 Myr, respectively), which are significantly different to those obtained when assuming that the stars are coeval. However, we notice that the model-derived cooling age of the binary system (Methods 1 and 2) is equal to the mean cooling age of the model-derived individual masses (Method 3).

The difference in the estimated cooling age of both stars ($\sim$ 63 Myr) is significant.
The model-derived cooling ages of the individual stars suggest that LP~349$-$25A is more evolved than LP~349$-$25B, which is consistent with LP~349$-$25A being an UCD and LP~349$-$25B being a brown dwarf. Even if both stars in a binary system were formed simultaneously, the star with a higher mass would evolve faster that the star with a lower mass. In the case of LP~349$-$25AB, the dynamical mass of the primary star LP~349$-$25A is consistent with an ultra-cool dwarf of 85.71 Jupiter masses, located above the hydrogen burning limit of about 78.5 M$_{J}$   \citep[][]{Chabrier23}, while the secondary star LP~349$-$25B has a dynamical mass of 67.11 Jupiter masses, which is bellow the hydrogen burning limit. Under these conditions, it is expected that LP~349$-$25A evolves faster than LP~349$-$25B.

\begin{figure*}
   \centering
     \plottwo{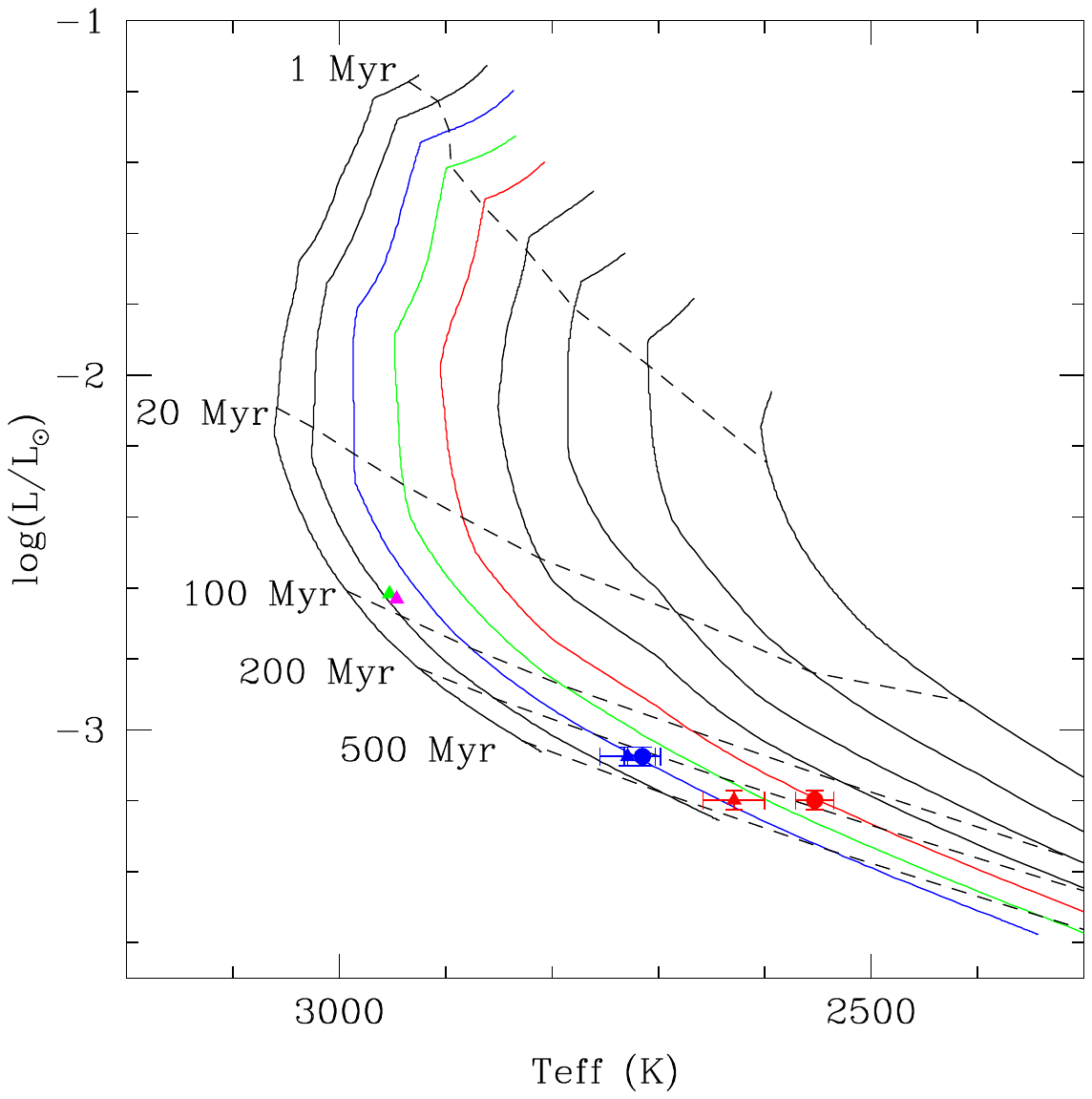}{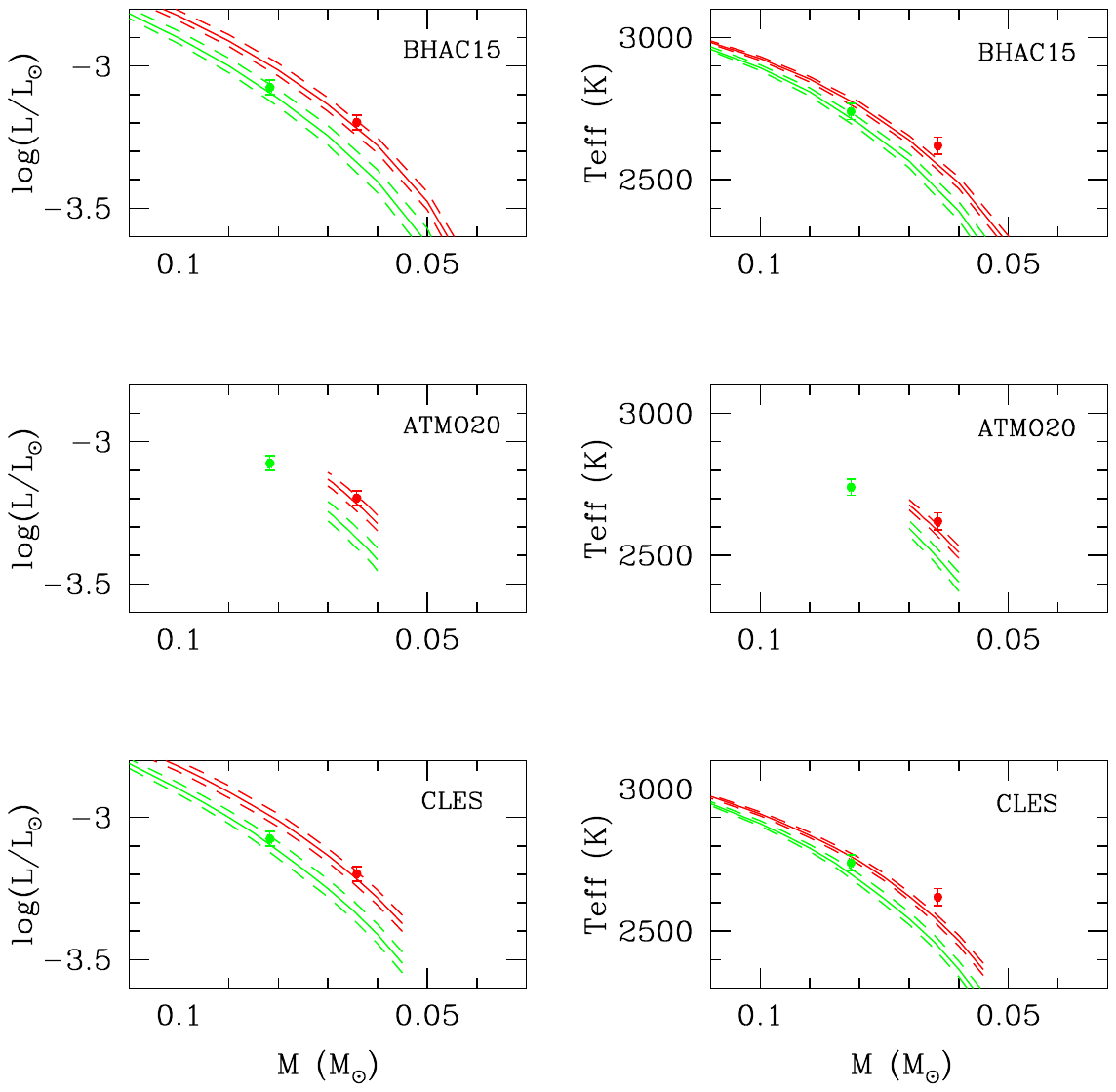}
    \caption{{\bf Left:} H-R diagram showing the luminosity of the LP~349$-$25A and LP~349$-$25B stars and the model-derived $T_{eff}$ determined from Method 3  (blue and red filled circles). 
Blue and red filled triangles correspond to the luminosity and $T_{eff}$  of both stars obtained by \citet[][]{Dupuy17}. For comparison, green and magenta data points correspond to the luminosity the stars in the younger binary M7 LSPM~J1314$+$1320AB \citep[][]{Dupuy16} and the model-derived $T_{eff}$ determined from Method 3. BHAC15 isomass tracks are shown in steps of 0.01 $M_{\odot}$ with the 0.06, 0.07 and 0.08 $M_{\odot}$ tracks highlighted in blue, green and red. Isochrones from 1 Myr to 500 Myr are indicated by dash lines. 
{\bf Right:} Comparison of our measured individual masses of both stars and the empirically derived temperature and luminosity by \citet[][]{Dupuy17} with those derived from the Stellar Evolutionary Models. The mass, luminosity and $T_{eff}$ of the main (green) and secondary (red) stars are plotted as filled circles with 1$\sigma$ error bars. The solid lines correspond to the Model-derived Isochrones that provide the best fit (see Table~\ref{tab_6}).
The dash lines show the isochrones of the best fits adding $\pm$1$\sigma$.
}
    \label{fig_HR}%
    \end{figure*}

Pre-main-sequence stellar models are commonly used to infer masses by placing objects on the H-R diagram. To test the accuracy of the masses derived from models, we use the effective temperatures (2729$^{+26}_{-27}$ and 2629$^{+29}_{-27}$ K)  and luminosities (log($L_{bol}/L_{\odot}$) = $-$3.075$\pm0.026$ dex  and $-$3.198$\pm0.027$ dex) of LP~349$-$25A and B obtained by  \citet[][]{Dupuy17} to derive mass and age (see Tables~\ref{tab_5} and \ref{tab_6}). 
Figure~\ref{fig_HR}  shows the estimated luminosity and temperatures of both stars compared to BHAC15 evolutionary model tracks. We also include in this figure the estimated values that we obtain from Model 3 (see Tables~\ref{tab_5} and \ref{tab_6}).

The relatively small, but significant, discrepancy between the H$-$R diagram derived mass (0.0777 and 0.0683 M$_\odot$) and our dynamical masses (0.0819 and 0.0641 M$_\odot$) suggest relatively small errors in the spectral type$-T_{eff}$ relations, which are calibrated using BT-Settl model atmospheres, systematic errors in the evolutionary models, or some combination of both. 
There is a significant difference between the H$-$R diagram derived ages (230 Myr) and the ages (262$\pm$21 and 199$\pm$12 Myr) that we obtain using Model 3 (see Table~\ref{tab_5}). 
This difference is due to the coeval age of both stars assumption in the estimates using the H$-$R diagram.
Regardless  the cause of the discrepancy, this test case shows that masses derived from the H$-$R diagram can harbor large systematic errors.

These results suggest that with a model-derived estimated cooling age of about 262 Myr for the binary system, both stars should show different age characteristics. 
For instance, the model-derived Li/Li$_{init}$ of the individual stars suggest that LP~349$-$25A has exhausted all its original Lithium, while LP~349$-$25B may still have a very small fraction ($\sim$ 0.7$\%$) of its original Lithium. Thus, a future search for Lithium absorption in the individual stars might show that  LP~349$-$25B still has some Lithium remaining.

Such a young age implies that LP~349$-$25AB is a pair of pre-main-sequence stars with masses of 85.71$\pm$0.64 M$_{Jup}$ and 67.11$\pm$0.51 M$_{Jup}$. 
At a distance of only 14.122$\pm$0.057 pc, this is the nearest pre-main-sequence binary system containing very low mass stars ($<$0.085 M$_{\odot}$). Furthermore, this the nearest binary system composed by an UCD and a brown dwarf.

The estimated dynamical masses of both stars, together with the luminosities and effective temperatures obtained by \citet[][]{Dupuy17}, can be used to test stellar evolutionary models.
Figure~\ref{fig_HR} shows a comparison of these measured and empirically derived quantities with those derived from the stellar evolutionary models BHAC15, ATMO20-CEQ and CLES-solar. To make a direct comparison we plot the Model-derived isochrones that provide the best fit (see Table~\ref{tab_6}). This figure shows that the evolutionary models BHAC15 and CLES-solar reproduce quite well the estimated mass, luminosities and effective temperature for the UCD LP~349$-$25A, but fails in the case of the brown dwarf LP~349$-$25B. On the other hand, ATMO20-CEQ reproduces well the observed quantities for the brown dwarf LP~349$-$25B, however, this particular family of models with equilibrium chemistry have a grid sample of masses adequate only for substellar objects (0.0005 $M_{\odot}$ to 0.075 $M_{\odot}$) and do not cover higher stellar masses, such as  that of LP~349$-$25A. 

\subsection{The Nature of LP~349$-$25AB} \label{sec:nature}

LP~349$-$25AB is a binary system that was found to comprise and M8 and an M9 ultra-cool dwarfs with bolometric luminosities $log (L_{bol}/L_{\odot})$ = $-$3.075$\pm$0.026 dex and $-$3.198$\pm$0.027 dex \citep[][]{Dupuy17}.
It is somewhat surprising that  LP~349$-$25AB has turned out to be binary system, where the stars have dynamical masses consistent with the main star being a young ultra-cool dwarf and the secondary star a brown dwarf (see Table~\ref{tab_3}).
The BHAC15 models predict a cooling age of 262 $\pm$ 21 Myr for LP~349$-$25A and 199 $\pm$ 12 Myr for LP~349$-$25B. 
In addition, using these evolutionary models, we find that Lithium in LP~349$-$25A is expected to be completely depleted, and that the remaining lithium fraction of LP~349$-$25B is about 0.7$\%$. These depleted fractions are consistent with the no detection of Lithium absorption in integrated light \citep[][]{Reiners09}.
These results suggest that LP~349$-$25AB is a pair of pre-main sequence stars with different model-derived cooling ages, where the secondary star is less evolved than the primary star.
In addition, although LP~349$-$25B has a dynamical mass consistent with being a brown dwarf, the estimated spectral types and luminosities of both stars are consistent with both stars being very low mass M-dwarfs, a consequence of both being pre-main sequence stars. 

\section{Conclusions and Final Remarks}  \label{sec:conclusions}

LP~349$-$25AB is an unusual ultra-cool dwarf binary system.
The radio continuum emission of the stellar components is quiescent, with no evidence of circular polarization, flaring or variability.
VLBA observations of the late M8$-$M9 dwarf binary system LP~349$-$25AB obtained at 11 epochs over 2020-2021 reveal that both components are radio emitters, with the componente LP~349$-$25A being the dominant radio emitter in all epochs. No circular polarization, nor outbursts were observed  from both sources. The primary star presents a small temporal flux density variation over a time span of months, while the secondary star does not show radio flux density variation. 

This binary system is one of the few UCD systems observed with multiple radio-emitting components. LP~349$-$25AB is only the second multiple UCD system probed with VLBI after the much older L dwarf 2M~J0746$+$2000AB binary system (4.4$-$5.1 Gyr) \citep[][]{Dupuy17,Zhang20}. The younger M7 LSPM~J1314$+$1320AB binary system (80.8 $\pm$ 2.5 Myr) has also been observed with VLBI, but only one component was found to be radio emitter \citep[][]{Dupuy16}.

Combining precise VLBI astrometry observations with optical/IR relative astrometric observations enables a precise measurement of the mass ratio of the two components, and thus their individual masses. The combined astrometric fit gives masses of 0.08188 $\pm$ 0.00061 $M_{\odot}$ and 0.06411 $\pm$ 0.00049 $M_{\odot}$ for LP~349$-$25A and LP~349$-$25B, respectively, indicating that the primary star is an UCD, and that the secondary component does not exceed the minimum stellar mass threshold. These measurements represent the most precise individual mass estimates of UCDs to date, which follows from the precise high spatial resolution of VLBI imagery together with precise relative astrometry extending nearly two decades, which covers more than one orbital period of the system.

We have used the estimated dynamical masses of both stars, together with the estimated luminosities and effective temperatures of both stars, to test the BHAC15, ATMO20-CEQ and CELES-solar stellar evolutionary models. We find that BHAC15 and CELES-solar reproduce quite well the observed parameters of the higher mass star LP~349$-$25A, however, they fail to reproduce the observed parameters of the lower mass star LP~349$-$25B. On the other hand, ATMO20-CEQ, which only contains a grid sample of masses adequate for substellar objects, reproduces quite well the observed parameters of the lower mass star LP~349$-$25B.

Using stellar evolutionary tracks we find that LP~349$-$25AB has a cooling age of 262 Myr.
Furthermore, we also find that the model-derived cooling age of  LP~349$-$25A is 262 Myr, while the model-derived cooling age of  LP~349$-$25B is 198 Myr.
These different cooling ages suggest that the secondary star LP~349$-$25B is less evolved than the primary star LP~349$-$25A. This result is consistent with the main star being an UCD, and the secondary star being a brown dwarf with a mass bellow the expected mass limit of hydrogen burning
\citep[$\sim$78.5 M$_{J}$;][]{Chabrier23}.

Such a young age implies that LP~349$-$25AB is a pair of pre-main-sequence stars with masses of 85.71 $\pm$ 0.64 M$_{Jup}$ and 67.11 $\pm$ 0.51 M$_{Jup}$, and that at a distance of only 14.122 $\pm$ 0.057 pc, this is the nearest pre-main-sequence binary system containing very low mass stars ($<$ 0.085 M$_{\odot}$) with direct mass measurements. 

Our results demonstrate that astrometric observations have the potential to fully characterize the orbital motions of binary and multiple stellar systems, and that precise stellar parameters of each star can be obtained by using stellar evolutionary models.

%-----------------------------------------------------------------

~~
~~
~~

\begin{acknowledgements}
\noindent
The authors would like to thank the anonymous referee for providing very useful comments that improved this paper.
We thank Jes\'us Hern\'andez and Carlos Rom\'an-Z\'uniga for valuable discussions about stellar evolutionary models.
S.C. acknowledges financial support from UNAM, and CONACyT, M\'exico. 
S.C. acknowledges financial support from the CONAHCyT project number CF-2023-I-232. 
The authors acknowledge support from the UNAM-PAPIIT  IN104521 and IN107324 grants.
The observations were carried out with the Very Long Baseline Array (VLBA), which is part of the National Radio Astronomy Observatory (NRAO). The NRAO is a facility of the National Science 
Foundation operated under cooperative agreement by Associated Universities, Inc.
This publication makes use of the SIMBAD database operated at the CDS, Strasbourg, France.
This work has made use of data from the European Space Agency (ESA) mission Gaia ({https://www.cosmos.esa.int/gaia}), processed by the Gaia Data Processing and Analysis Consortium (DPAC, {https://www.cosmos.esa.int/web/gaia/dpac/consortium}). Funding for the DPAC has been provided by national institutions, in particular the institutions participating in the Gaia Multilateral Agreement.
\end{acknowledgements}

\vspace{5mm}
\facilities{VLBA}

\software{AIPS \citep[][]{Greisen03}, 
               astropy \citep{Astropycol13,Astropycol18},
               corner\citep{Foremanmackey16}, 
               emcee \citep{Foremanmackey13}, 
               lmfit \citep[][]{Newville20}, 
               scipy \citep[][]{Newville20},
               matplotlib,\citep{Hunter07},
               numpy\citep{Vanderwalt11}. 
          }

~~
~~
~~

{~~~~~~~~~~~~~~~~~~~~~{\bf ORCID iDs}}

Salvador Curiel {https:/orcid.org/0000-0003-4576-0436}

Gisela N. Ortiz-Le\'on {https:/orcid.org/0000-0002-2863-676X}

Amy J. Mioduszewski {https:/orcid.org/0000-0002-2564-3104}

~~
~~
~~

{}

%\clearpage

%%%%%%%%%%%%%%%%%%%%%%%%%%%%%%%%%%%%%%%%%%%%%%%%%%%%%

%--------------------------------------------------------------------
% Table 1

%\clearpage
\begin{deluxetable*}{cccccccc}
\tablecaption{Properties of the VLBA detections. \label{tab_1} }
\tablewidth{0pt}
\tablehead{
\colhead{Julian Date} & \colhead{$\alpha(J2000)$} & \colhead{$\sigma_\alpha$} & \colhead{$\delta(J2000)$} 
&  \colhead{$\sigma_\delta$}  & \colhead{Integrated flux density} & \colhead{rms}  \\
& \colhead{(h:m:s)}& \colhead{(s)} & \colhead{($^{\rm o}$:$'$:$''$)} & \colhead{($''$)} & \colhead{(mJy)} 
& \colhead{(mJy~beam$^{-1}$)} \\
\colhead{(1)}  & \colhead{(2)}  & \colhead{(3)}  & \colhead{(4)}  & \colhead{(5)}  & \colhead{(6)}  & \colhead{(7)} 
}  
\startdata
\sidehead{LP349--25A}
\hline
2458902.25601 & 0:27:56.57629060 &  0.00000944 & 22:19:28.8231491 &  0.0001386 & 0.244$\pm$0.039 & 0.019 \\  
2458927.32128 & 0:27:56.58037892 &  0.00000790 & 22:19:28.8198101 &  0.0001150 & 0.203$\pm$0.031 & 0.016 \\  
2459259.40179 & 0:27:56.60691755 &  0.00000979 & 22:19:28.6865557 &  0.0001441 & 0.127$\pm$0.024 & 0.010 \\  
2459287.32604 & 0:27:56.61129332 &  0.00000873 & 22:19:28.6823381 &  0.0001277 & 0.147$\pm$0.023 & 0.013 \\  
2459309.27251 & 0:27:56.61492416 &  0.00000621 & 22:19:28.6836204 &  0.0000887 & 0.206$\pm$0.022 & 0.012 \\  
2459333.19681 & 0:27:56.61880737 &  0.00000599 & 22:19:28.6884587 &  0.0000852 & 0.204$\pm$0.021 & 0.011 \\  
2459339.17667 & 0:27:56.61970519 &  0.00000515 & 22:19:28.6897750 &  0.0000718 & 0.225$\pm$0.020 & 0.011 \\  
2459373.09066 & 0:27:56.62426640 &  0.00000724 & 22:19:28.6979596 &  0.0001047 & 0.158$\pm$0.021 & 0.012 \\  
2459447.88609 & 0:27:56.62915682 &  0.00000720 & 22:19:28.6900412 &  0.0001042 & 0.196$\pm$0.025 & 0.011 \\  
2459492.76322 & 0:27:56.62955737 &  0.00000611 & 22:19:28.6605022 &  0.0000871 & 0.231$\pm$0.024 & 0.011 \\  
2459570.55023 & 0:27:56.63183259 &  0.00000965 & 22:19:28.5889998 &  0.0001419 & 0.245$\pm$0.039 & 0.015 \\  
\hline
\sidehead{LP349--25B}
\hline
2459287.32604 & 0:27:56.60664322 &  0.00001628 & 22:19:28.7146897 &  0.0002425 & 0.081$\pm$0.020 & 0.013 \\  
2459309.27251 & 0:27:56.60999054 &  0.00001116 & 22:19:28.7102135 &  0.0001650 & 0.110$\pm$0.023 & 0.012 \\  
2459333.19681 & 0:27:56.61359010 &  0.00001166 & 22:19:28.7092159 &  0.0001726 & 0.087$\pm$0.020 & 0.011 \\  
2459339.17667 & 0:27:56.61443005 &  0.00001114 & 22:19:28.7082492 &  0.0001646 & 0.122$\pm$0.024 & 0.011 \\  
2459373.09066 & 0:27:56.61861414 &  0.00001095 & 22:19:28.7078198 &  0.0001618 & 0.088$\pm$0.020 & 0.012 \\  
2459447.88609 & 0:27:56.62282165 &  0.00000918 & 22:19:28.6802288 &  0.0001347 & 0.115$\pm$0.021 & 0.011 \\  
2459492.76322 & 0:27:56.62287105 &  0.00000894 & 22:19:28.6387254 &  0.0001311 & 0.120$\pm$0.021 & 0.011 \\  
2459570.55023 & 0:27:56.62469621 &  0.00001178 & 22:19:28.5472898 &  0.0001743 & 0.143$\pm$0.031 & 0.015 \\ 
 \enddata
%\tablenotetext{b}{.}
\end{deluxetable*}

%\clearpage

%--------------------------------------------------------------------
% Table 2

%\clearpage
\startlongtable
\begin{deluxetable}{cccccc}
\tablecaption{Relative Astrometry\tablenotemark{a} \label{tab_2} }
\tablewidth{0pt}
\tablehead{
\colhead{Julian day} & \colhead{$\Delta\alpha(J2000)$} & \colhead{$\sigma_{\Delta\alpha}$} & \colhead{$\Delta\delta(J2000)$} &  \colhead{$\sigma_{\Delta\delta}$}  & \colhead{Reference\tablenotemark{b}}\\
& \colhead{(mas)} & \colhead{(mas)} & \colhead{(mas)} & \colhead{(mas)} &  \\
\colhead{(1)}  & \colhead{(2)}  & \colhead{(3)}  & \colhead{(4)}  & \colhead{(5)}  & \colhead{(6)} 
}  
\startdata
\hline
2453190.48  &  27.481   &  4.791 &  121.942  &   9.802 & 1 \\
2453275.48  & 13.225    & 1.545  & 106.180  &   9.924 & 1 \\
2454067.48 & -103.062   &  1.872  & -72.487   &  1.732 & 2 \\
2454067.48  & -103.604  &   2.470  &  -68.133  &   2.977 & 2 \\
2454067.48 &  -101.343  &   1.744  &  -69.703  &   1.406 & 2 \\
2454436.48 &  -68.389  &   3.787 &  -111.732   &  3.922 & 2  \\
2454482.48 &  -64.583  &   0.170  & -121.106   &  0.222 &  3 \\
2454617.48 &  -38.503  &   0.886 & -118.922   &  0.989 & 2 \\
2454648.48 &  -29.617  &   0.241 &  -110.997  &   0.249 &  3 \\
2454648.48 &  -29.930  &   0.166 &  -110.695  &   0.226 &  3 \\
2454648.48 &  -30.021  &   0.364 &  -110.494  &   0.586 &  3 \\
2454699.48 &  -18.175  &   0.092 &  -104.136  &   0.100 &  3 \\
2454719.48 &  -13.548  &   0.160 &  -101.268  &   0.140 &  3 \\
2454820.48 &   11.227   &  1.213 &  -84.255   &  1.004 & 2 \\
2454820.48 &   10.584   &  1.054 &  -88.368   &  3.973 & 2 \\
2454994.48 &   50.839   &  3.085 &  -42.088   &  2.558 & 2 \\
2455103.48 &   70.436  &   0.316 &  -10.401  &   0.739 &  3 \\
2455181.48 &   82.660  &   0.401 &   12.501  &   0.437 &   3 \\
2455339.48 &   97.584  &   0.179 &   56.499  &   0.178 &   3 \\
2455941.48 &   43.905  &   0.349  &  123.848  &   0.666 &  3 \\
 \enddata
%%%%%
%\tablecomments{comments}
%\tablenotetext{}{Notes }
\tablenotetext{a}{The parameters presented here correspond to the projected relative position of the secondary star in $\Delta\alpha$ and $\Delta\delta$ coordinates, using the separation $\rho$ and the position angle P.A. of each observed epoch.}
\tablenotetext{b}{(1) \citet[][]{Forveille05}, (2) \citet[][]{Konopacky10}, (3) \citep[][]{Dupuy17}, (4) This work.}
\end{deluxetable}

%\clearpage

%--------------------------------------------------------------------
% Table 3

\startlongtable
\begin{deluxetable}{lccc}
%\rotate
\centering
\tabletypesize{\scriptsize}
\tablewidth{0pt}
\tablecolumns{4}
\tablecaption{Combined Astrometry Fits\tablenotemark{a} \label{tab_3}}             % title of Table
\tablehead{
\\ \colhead{Parameter} &  
%\colhead{Relative\tablenotemark{b}} & 
\colhead{AGA\tablenotemark{b}} &
\colhead{Least squares\tablenotemark{c}} &
\colhead{MCMC\tablenotemark{d}}  \\  % table heading 
 & \colhead{(1)}  & \colhead{(2)}  & \colhead{(3)} 
}
\startdata
& Fitted Parameters & &   \\
 \hline                        % inserts single horizontal line
$\mu_{\alpha}$ (mas yr$^{-1}$)  &  408.68 $\pm$ 0.40 & 408.65 $\pm$ 0.57  &  408.64$^{+0.59}_{-0.56}$  \\
$\mu_{\delta}$ (mas yr$^{-1}$)  &  $-$170.77 $\pm$ 0.39  & $-$170.80 $\pm$ 0.75 &  $-$170.81$^{+0.77}_{-0.74}$ \\
$\Pi$ (mas)            &  70.81 $\pm$  0.29  & 70.811 $\pm$ 0.089   & 70.810$^{+0.092}_{-0.088}$  \\
$P$ (days)             &    2816.03  $\pm$ 1.12   & 2815.99 $\pm$ 1.79 & 2816.00$^{+1.81}_{-1.82}$  \\
$T_{0}$ (Julian day)\tablenotemark{e}  &  2,457,734.12 $\pm$ 0.92 & 2,457,733.82 $\pm$ 12.34 &   2,457,733.89$^{+12.42}_{-12.41}$   \\ 
$e$                        &  0.0537 $\pm$ 0.0022 & 0.0537 $\pm$ 0.0010 & 0.0537$^{+0.0010}_{-0.0010}$   \\
$\omega$$_{A}$ (deg)      &   257.93 $\pm$ 0.12  & 257.89 $\pm$ 1.55 & 257.90$^{+1.56}_{-1.56}$ \\
$\Omega$ (deg)         &  216.36 $\pm$ 0.15  & 216.36 $\pm$ 0.10 &  216.36$^{+0.10}_{-0.10}$  \\
$a_{A}$ (mas)            &  63.91 $\pm$ 0.17  & 63.96 $\pm$ 1.31  & 63.99$^{+1.29}_{-1.35}$   \\
$i$ (deg)                    &   117.81 $\pm$ 0.14  & 117.812 $\pm$  0.070  &  117.812$^{+0.071}_{-0.071}$  \\
$q$ (m$_{B}$/m$_{A}$)    &   0.7830 $\pm$ 0.0038  & 0.784 $\pm$  0.029 &  0.785$^{+0.029}_{-0.029}$  \\
$V_{0}$ (km~s$^{-1}$)     &   $-$8.24 $\pm$ 2.47  & $-$8.24 $\pm$  0.58 &  $-$8.24$^{+0.59}_{-0.58}$  \\
\hline                        % inserts single horizontal line
&   Other Parameters & &  \\
\hline                        % inserts single horizontal line
$D$ ~(pc)                          &  14.122 $\pm$ 0.057  & 14.122 $\pm$ 0.018 & 14.122 $\pm$ 0.012 \\
$a_{B}$ ~(mas)                &  81.61 $\pm$ 0.45 & 81.56  $\pm$   3.43 & 81.53  $\pm$ 2.33  \\
$a$ ~(mas)               &  145.52 $\pm$ 0.48  & 145.52 $\pm$ 3.68 & 145.52  $\pm$ 2.50  \\
$a_{A}$ ~(AU)           &  0.9024  $\pm$ 0.0044 & 0.903 $\pm$ 0.019 & 0.904 $\pm$ 0.013   \\
$a_{B}$ ~(AU)           & 1.1525  $\pm$ 0.0053 & 1.152 $\pm$ 0.049 & 1.151 $\pm$ 0.033  \\
$a$ ~(AU)        &  2.0550  $\pm$ 0.0069  & 2.055 $\pm$ 0.052 & 2.055  $\pm$ 0.035  \\
$m_{A}$ ~(M$_\odot$)      &  0.08188  $\pm$ 0.00061  & 0.0818 $\pm$ 0.0047 & 0.0818 $\pm$  0.0032 \\
$m_{B}$ ~(M$_\odot$)      &  0.06411  $\pm$ 0.00049  & 0.0642 $\pm$ 0.0044 & 0.0642 $\pm$ 0.0030 \\
$m$ ~(M$_\odot$)   &  0.14599  $\pm$ 0.00070  & 0.1460 $\pm$ 0.0064 & 0.1460 $\pm$ 0.0043  \\
$m_{A}$ ~(M$_{J}$)         &    85.71  $\pm$ 0.64  & 85.70 $\pm$ 4.90 & 85.67 $\pm$ 3.33 \\
$m_{B}$ ~(M$_{J}$)         &    67.11  $\pm$ 0.51  & 67.21 $\pm$  4.57 & 67.24  $\pm$ 3.11  \\
$m$ ~(M$_{J}$)       & 152.82  $\pm$ 0.73  & 152.91 $\pm$ 6.70 & 152.92 $\pm$  4.55\\
$\Delta$$\alpha$(A) ~(mas)\tablenotemark{f}   &  0.16 & 0.15 & 0.15  \\
$\Delta$$\delta$(A) ~(mas)\tablenotemark{f}    &  0.19 & 0.18 & 0.18  \\
$\Delta$$\alpha$(B) ~(mas)\tablenotemark{f}   &  0.23 & 0.19 & 0.19  \\
$\Delta$$\delta$(B) ~(mas)\tablenotemark{f}    &  0.26 & 0.20 & 0.20  \\
$\Delta$$\alpha$ ~(mas)\tablenotemark{f} &  1.43 & 1.86 & 1.86  \\
$\Delta$$\delta$ ~(mas)\tablenotemark{f}  &   2.92 &  3.38 & 3.38   \\
$\Delta$$V$(A) ~(km~s$^{-1}$)\tablenotemark{f} &   0.61 & 0.34 & 0.34  \\
$\Delta$$V$(B) ~(km~s$^{-1}$)\tablenotemark{f}  &  1.08 & 0.63 & 0.63   \\
$\chi^2$, $\chi^2_{red}$\tablenotemark{g}   & 155.45, 2.16 & 155.44, 2.16    & 155.45, 2.16  \\
\enddata
%%%%%
\tablenotetext{a}{The parameters presented here were obtained with the AGA, non-linear least squares and MCMC algorithms. The subindex A and B correspond to the main star (LP~349$-$25A) and the secondary star (LP~349$-$25B), respectively.}
\tablenotetext{b}{The AGA combined astrometric fit is obtained by fitting simultaneously 
the absolute astrometry of both stars, the relative astrometry of the binary system and the radial velocity of both stars (see text).  All the free parameters are fitted simultaneously.}
\tablenotetext{c}{Non-linear least squares combined fit of the absolute and relative astrometric data, and the radial velocity data of both stars.}
\tablenotetext{d}{MCMC combined fit of the absolute and relative astrometric data, and the radial velocity data of both stars.}
\tablenotetext{e}{Time of the periastron passage.}
\tablenotetext{f}{RMS dispersion of the residuals. The first two terms correspond to the rms residuals of the absolute astrometry of the primary, the next two to the rms residuals of the absolute astrometry of the secondary, the next two to the rms residuals of the relative astrometry from the literature, and the last two are the rms residuals of the RVs.}
\tablenotetext{g}{$\chi^2$ and reduced $\chi^2$ of the astrometric fit. In all cases the residuals of the relative astrometry dominates the residuals of the fit.}
\end{deluxetable}

%\clearpage

%--------------------------------------------------------------------
% Table 4

\startlongtable
\begin{deluxetable}{lcccc}
%\rotate
\centering
\tabletypesize{\scriptsize}
\tablewidth{0pt}
\tablecolumns{5}
\tablecaption{Comparison with Previous Orbital Fits\tablenotemark{a} \label{tab_4}}             % title of Table
\tablehead{
\colhead{Parameter} &  
\colhead{This work} & 
\colhead{\citet[][]{Konopacky10}} &
\colhead{\citet[][]{Dupuy10}} &
\colhead{\citet[][]{Dupuy17}} \\
\colhead{} & \colhead{(1)} & \colhead{(2)} & \colhead{(3)} & \colhead{(4)}
}
\startdata
$\mu_{\alpha}$ (mas yr$^{-1}$)  & 408.68 $\pm$ 0.40 & ... &  ...  & 407.9 $_{-1.7}^{+1.7}$  \\
$\mu_{\delta}$ (mas yr$^{-1}$)   & $-$170.77 $\pm$ 0.39 & ... &  ...  & $-$170.4 $_{-1.3}^{+1.3}$  \\
$\Pi$ (mas)\tablenotemark{e}  &  70.81 $\pm$  0.29  & 75.82 (fixed) &   75.8 $\pm$ 1.6 (fixed)  & 69.2$_{-0.9}^{+0.9}$    \\
$P$ (days)            &  2816.03  $\pm$ 1.12 & 2670 $\pm$ 135 &  2834.34$_{-14.61}^{+14.61}$   & 2811.69$_{-5.11}^{+5.11}$  \\
$T_{0}$ (Julian day)  &  2,457,734.12 $\pm$ 0.92 & 2002.5 $\pm$ 0.8 &  2,454,860.5$_{-24}^{+26}$  & 2,457,758$_{-14}^{+15}$   \\ 
$e$                         &  0.0537 $\pm$ 0.0022  & 0.08 $\pm$ 0.02 & 0.051$_{-0.003}^{+0.003}$ & 0.0468$_{-0.0018}^{+0.0019}$   \\
$\omega_{A}$ (deg)     & 257.93 $\pm$ 0.12 & 289 &   70  & 82.2   \\
$\omega_{B}$ (deg)     & 77.93 & 109$_{-22}^{+37}$ &   250$_{-3}^{+4}$ & 262.2$_{-1.8}^{+1.8}$   \\
$\Omega$ (deg)    & 216.36 $\pm$ 0.15  & 213.8 $\pm$ 1.1 &  35.95$_{-0.12}^{+0.12}$ & 36.64$_{-0.10}^{+0.10}$  \\
$a_{A}$ (mas)              &  63.91 $\pm$ 0.17  & ... &  ...  & ...    \\
$a$ (mas)              &  145.52 $\pm$ 0.48 & 141 $\pm$ 7 &  146.7$_{-0.6}^{+0.6}$  & 145.99$_{-0.18}^{+0.17}$    \\
$i$ (deg)                & 117.81 $\pm$ 0.14  & 118.7 $\pm$ 1.5 &  117.24$_{-0.14}^{+0.14}$  &  117.36$_{-0.10}^{+0.11}$  \\
$q$ ~(m$_{B}$/m$_{A}$)         & 0.7830 $\pm$ 0.0038  & ... & ... & ...  \\
$V_{0}$~(km~s$^{-1}$)         & $-$8.24 $\pm$ 2.47  & ... & ... & ...  \\
$m  ~(M_\odot)$   &  0.14599 $\pm$ 0.00070 & 0.121 $\pm$ 0.009 &  0.120$_{-0.007}^{+0.008}$  & 0.158$_{-0.007}^{+0.006}$  \\
$D$ ~(pc)                                & 14.122 $\pm$ 0.057  & 13.19 $\pm$ 0.28 (fixed) &  ... &14.45$_{-0.19}^{+0.18}$  \\
$\chi^2$, $\chi^2_{red}$    & 155.45, 2.16  & 2.15 & ...  & ... \\
%
%\hline                        % inserts single horizontal line
\enddata
%%%%%
\tablenotetext{a}{The parameters presented in column (1) were obtained with the AGA code (see column (1) in Table \ref{tab_3}). The other three columns show the fitted parameters obtained by 
\citet[][]{Konopacky10},
\citet[][]{Dupuy10}, and
\citet[][]{Dupuy17}.
The subindex A and B correspond to the main star (LP~349$-$25A) and the secondary star (LP~349$-$25B), respectively.
The residuals of the relative astrometry are substantially larger than the residuals of the absolute fit and thus it is the main source of $\chi^2$.}
\end{deluxetable}

\clearpage

%---------------------------------------------------------------
% Table 5
%
\startlongtable
\begin{deluxetable}{lccc}
%\rotate
\centering                          % used for centering table
\tabletypesize{\scriptsize}
\tablewidth{0pt}
\tablecolumns{2}
\tablecaption{Observed and Model-derived Properties \label{tab_5}} 
\tablehead{                % inserts double horizontal lines
Parameters & Method 1 & Method 2 & Method 3       % \\    % table heading 
}
\decimals
\startdata
\hline                        % inserts single horizontal line
&  & Input Parameters &   \\
\hline                        % inserts single horizontal line
m (M$\odot$)            &   0.1460   &  0.1460 $\pm$ 0.0007 & 0.1460 $\pm$ 0.0007 \\
m$_{A}$ (M$\odot$)                      &   ...           &  0.0819 $\pm$ 0.0006 & 0.0819 $\pm$ 0.0006  \\
m$_{B}$ (M$\odot$)                      &   ...           &  0.0641 $\pm$ 0.0005 &  0.0641 $\pm$ 0.0005 \\
Mass ratio q                          &   ...           &   0.7830 $\pm$ 0.0038 & 0.7830 $\pm$ 0.0038 \\
log(L$_{bol}$(AB)) [L$_\odot$]     &  ... &  $-$2.851$^{+0.026}_{-0.027}$ & ... \\
log(L$_{bol}$(A)) [L$_\odot$]   &  $-$3.075 $\pm$ 0.026  &  ... & $-$3.075 $\pm$ 0.026 \\
log(L$_{bol}$(B)) [L$_\odot$]   &  $-$3.198 $\pm$ 0.027 &  ...  & $-$3.198 $\pm$ 0.027\\
\hline
& & Derived from BHAC15   & \\
\hline
m$_{A}$  (M$\odot$)                          &  0.0777   & ... & ... \\
m$_{B}$  (M$\odot$)                          &  0.0683   & ... & ... \\
log(L$_{bol}$(A)) [L$_\odot$]      &  ...           & $-$3.050$^{+0.025}_{-0.025}$ & ... \\
log(L$_{bol}$(B)) [L$_\odot$]      &  ...           & $-$3.286$^{+0.030}_{-0.030}$ & ... \\
Mass ratio q                              &  0.88  &   ... & ... \\
Age$_{A}$ (Myr) & 230 $\pm$ 16 &   232 $\pm$ 16  & 262 $\pm$ 21  \\
Age$_{B}$ (Myr) & 230 $\pm$ 16 &   232 $\pm$ 16  & 199 $\pm$ 12  \\
$T_{eff}$(A) ($K$) &  2699 $\pm$ 17  & 2745$^{+14}_{-15}$ &  2716 $\pm$ 18     \\
$T_{eff}$(B) ($K$) &  2574 $\pm$ 19  & 2502$^{+21}_{-23}$ & 2553 $\pm$ 18   \\
R$_{A}$ ($R_{\odot}$) &  0.1297$^{+0.0026}_{-0.0022}$  & 0.1322$^{+0.0026}_{-0.0021}$ & 0.128$^{+0.002}_{-0.002}$   \\
R$_{B}$ ($R_{\odot}$) &  0.1240$^{+0.0019}_{-0.0022}$  & 0.1213$^{+0.0018}_{-0.0023}$  & 0.126$^{+0.003}_{-0.002}$  \\
log($g_{A}$) [$cm s^{-2}$] &  5.101$^{+0.014}_{-0.015}$  & 5.107 $\pm$ 0.015 & 5.135$^{+0.016}_{-0.017}$     \\
log($g_{B}$) [$cm s^{-2}$] &  5.085$^{+0.014}_{-0.015}$  & 5.076 $\pm$ 0.014 &  5.043$^{+0.013}_{-0.014}$    \\
$Li_{A}/Li_{ini}$ &  $<$0.0001 & $<$0.0001  &  $<$0.0001  \\
$Li_{B}Li_{ini}$  &  $<$0.0001 & 0.003 $\pm$ 0.001  &  0.007 $\pm$ 0.003   \\
\hline
& & Derived from CLES-solar   & \\
\hline
log(L$_{bol}$(A)) [L$_\odot$]      &  ...           & $-$3.049$^{+0.024}_{-0.025}$ & ... \\
log(L$_{bol}$(B)) [L$_\odot$]      &  ...           & $-$3.287$^{+0.029}_{-0.030}$ & ... \\
Age (A) (Myr) & ... &   224 $\pm$ 16  & 255 $\pm$ 19  \\
Age (B) (Myr) & ... &   224 $\pm$ 16  & 191 $\pm$ 12  \\
$T_{eff}$(A) ($K$) &  ... & 2733 $\pm$ 15 &  2705$^{+16}_{-18}$     \\
$T_{eff}$(B) ($K$) &  ...  & 2486$^{+21}_{-22}$ & 2536 $\pm$ 19    \\
R$_{A}$ ($R_{\odot}$) &  ...  & 0.1334$^{+0.0023}_{-0.0024}$ & 0.1292$^{+0.0025}_{-0.0022}$   \\
R$_{B}$ ($R_{\odot}$) &  ...  & 0.1227$^{+0.0020}_{-0.0020}$  & 0.1276 $\pm$ 0.0020  \\
log($g_{A}$) [$cm s^{-2}$] &  ...  & 5.101 $\pm$ 0.016 & 5.129$^{+0.015}_{-0.017}$     \\
log($g_{B}$) [$cm s^{-2}$] &  ...  & 5.068 $\pm$ 0.014 &  5.034$^{+0.013}_{-0.014}$    \\
$Li_{A}/Li_{ini}$ &  ... & $<$0.0001  &  $<$0.0001  \\
$Li_{B}Li_{ini}$  &  ... & 0.0050 $\pm$ 0.003 &  0.025$^{+0.019}_{-0.016}$   \\
\hline
& & Derived from ATMO20-CEQ   & \\
\hline
Age (B) (Myr)                      & ...   & ...  & 198    $\pm$ 12  \\
$T_{eff}$(B) ($K$)              &  ...  & ...  & 2585     $\pm$ 20    \\
R$_{B}$ ($R_{\odot}$)       &  ...  & ...  &  0.1229 $\pm$ 0.0018  \\
log($g_{B}$) [$cm s^{-2}$] &  ...  & ...  &  5.066   $\pm$ 0.013    \\
\enddata
\end{deluxetable}

\clearpage

%---------------------------------------------------------------
% Table 6
%
\startlongtable
\begin{deluxetable}{lccc}
%\rotate
\centering                          % used for centering table
\tabletypesize{\scriptsize}
\tablewidth{0pt}
\tablecolumns{2}
\tablecaption{Comparizon of Model-derived Stellar Properties obtained with Method 3 \label{tab_6}} 
\tablehead{                % inserts double horizontal lines
\colhead{Parameters\tablenotemark{a}} & \colhead{BHAC15} & 
\colhead{CLES-solar} & \colhead{ATMO20-CEQ\tablenotemark{b}}
}
\decimals
\startdata
\hline
Age (A) (Myr)                      & 262 $\pm$ 21  & 255 $\pm$ 19  & 262 $\pm$ 21  \\
$T_{eff}$(A) ($K$)              &  2716 $\pm$18  & 2705$^{+16}_{-18}$  & 2716 $\pm$ 18    \\
R$_{A}$ ($R_{\odot}$)       & 0.128 $\pm$ 0.002 & 0.1292$^{+0.0025}_{-0.0022}$ &  0.128 $\pm$ 0.002 \\
log($g_{A}$) [$cm s^{-2}$] &  5.135$^{+0.016}_{-0.017}$  & 5.129$^{+0.015}_{-0.017}$ &  5.135$^{+0.016}_{-0.017}$    \\
$Li_{A}Li_{ini}$           &  $<$0.0001 & $<$0.0001   &  $<$0.0001   \\
\hline
Age (B) (Myr)                      & 199 $\pm$ 12  & 191 $\pm$ 12 & 198 $\pm$ 12  \\
$T_{eff}$(B) ($K$)              &  2553 $\pm$ 18  & 2536 $\pm$ 19 & 2585 $\pm$ 20    \\
R$_{B}$ ($R_{\odot}$)       & 0.126$^{+0.003}_{-0.002}$  & 0.1276 $\pm$ 0.0020 &  0.1229 $\pm$ 0.0018  \\
log($g_{B}$) [$cm s^{-2}$] &  5.043$^{+0.013}_{-0.014}$  & 5.034$^{+0.013}_{-0.014}$ &  5.066   $\pm$ 0.013    \\
$Li_{B}Li_{ini}$          &  0.006 $\pm$ 0.003 & 0.025$^{+0.019}_{-0.016}$   &  ...   \\
\enddata
\tablenotetext{a}{Sun-index A corresponds to LP~349$-$25A and sub-index B corresponds to  LP~349$-$25B.}
\tablenotetext{b}{The properties of LP~349$-$25A and  LP~349$-$25B were obtained using BHAC15 and ATMO20-CEQ, respectively.}
\end{deluxetable}

%--------------------------------------------------------------------
% Table 7

%\clearpage
\startlongtable
\begin{deluxetable}{lccccccc}
\tablecaption{Observed and Model-derived Properties. \label{tab_7}}
\tablewidth{0pt}
\tablehead{
\colhead{ } & \colhead{ } & \colhead{This work} & & 
\colhead{ } &  \colhead{Dupuy \& Liu (2017)} & \\
\colhead{Property} & 
\colhead{Primary} & \colhead{Combined} & \colhead{Secondary} & 
\colhead{Primary} & \colhead{Combined} & \colhead{Secondary}
}  
\startdata
%\hline
m (M$\odot$) &  ... & 0.14622$\pm$0.00083 & ... & ... & 0.158$^{0.006}_{0.007}$ & ...  \\
m$_{A}$ (M$\odot$) & 0.08169$\pm$0.00036   & ... & 0.0642$\pm$0.0003 & ... &  ... & ...   \\
m$_{B}$ (M$\odot$) & 0.06425$\pm$0.00029   & ... & 0.0642$\pm$0.0003 & ... & ...  & ...   \\
Mass ratio q                       & ... &  0.7855$\pm$0.0022 & ... &   ... & ... & ...  \\
log(L$_{bol}$) [L$_\odot$]\tablenotemark{a} &  ... & $-$2.831$^{+0.026}_{-0.027}$ & ...  & $-$3.075$^{+0.026}_{-0.027}$  &  ... & $-$3.198$^{+0.026}_{-0.027}$ \\
\hline
& Derived & from & Baraffe et al. (2015) & Evolutionary & Models &  & \\
\hline
m (Jupiter) & ... & ... & ...  & 89$\pm$4 & ... &  78$^{+3}_{-4}$  \\
log(L$_{bol}$) [L$_\odot$] &  $-$3.032$^{+0.02}_{-0.03}$  & ... & $-$3.262 $^{+0.03}_{-0.03}$   & $-$3.074$^{+0.029}_{-0.032}$  & ... & $-$3.20$\pm$0.03  \\
Mass ratio q &  ... & ... &   ... &  ... & 0.88$^{+0.03}_{-0.04}$ & ...    \\
Age $t$ (Gyr) &  ... & 0.220$^{+0.000}_{-0.000}$ & ...  & ... & 0.271$^{+0.022}_{-0.029}$ &  ...   \\
$T_{eff}$ ($K$) &  2754 $\pm$ 15  &  ... & 2521$^{+21}_{-23}$   & 2740$^{+32}_{-29}$  & ... & 2620 $\pm$ 30  \\
Radius ($R_{\odot}$) &  0.134 $\pm$ 0.003  & ... &  0.123$^{+0.002}_{-0.003}$   & 1.255$^{+0.019}_{-0.024}$  & ... & 1.193$^{+0.020}_{-0.017}$   \\
log($g$) [$cm s^{-2}$] &  5.143$^{+0.02}_{-0.03}$  & ... & $-$3.262 $^{+0.03}_{-0.03}$   & 5.143$^{+0.024}_{-0.020}$  & ... & 5.133$\pm$0.024  \\
log($Li/Li_{ini}$)  &  $<$$-$4.0 & ... & $-$2.44$^{+0.3}_{-0.3}$   & $<$$-$4.0  & ... & $<$$-$4.0  \\
 \enddata
\tablenotetext{b}{Combined luminosity and individual luminosity of each star obtained by \citet[][]{Dupuy17}.}
\end{deluxetable}
%\clearpage

%-------------------------------------- Two column figure (place early!)
% FIGURE 1

%-------------------------------------- Two column figure (place early!)
% FIGURE 2

%-------------------------------------- Two column figure (place early!)
% FIGURE 3
 
%-------------------------------------- Two column figure (place early!)
% FIGURE 4

%-------------------------------------- Two column figure (place early!)
% FIGURE 5
%

%-------------------------------------- Two column figure (place early!)
% FIGURE 6

%-------------------------------------- Two column figure (place early!)
% FIGURE 7


\begin{thebibliography}{}
\bibitem[Astropy Collaboration et al.(2013)]{Astropycol13} Astropy Colaboration, Robitaille, T. P., Tollerun, E. J., et al., \aap, 558, A33 (2013)
\bibitem[Astropy Collaboration et al.(2018)]{Astropycol18} Astropy Colaboration, Price-Whelan, A. M., SipH ocz, M. M., et al., \aj, 156, 123 (2018)
\bibitem[Baraffe et al.(2015)]{Baraffe15} Baraffe, I., Homeier, D., Allard, F., \& Chabrier, G. 2015, \aap, 577, A42. 
\bibitem[Berger et al.(2001)]{Berger01} Berger, E., et al. 2001, Nature, 410, 338. 
\bibitem[Berger(2002)]{Berger02} Berger, E. 2002, \apj, 572, 503. 
\bibitem[Berger(2006)]{Berger06} Berger, E. 2006, \apj, 648, 629. 
\bibitem[Berger et al.(2009)]{Berger09} Berger, E., et al. 2009, \apj, 695, 310. 
\bibitem[Blake et al.(2010)]{Blake10} Blake, C. H., et al. 2010, \apj, 723, 684. 
\bibitem[Bond et al.(2004)]{Bond04} Bond, I. A., et al. 2004, \apjl, 606, L155. 
\bibitem[Boss(2006)]{Boss06} Boss, A.P. 2006, \apj, 643, 501. 
\bibitem[Bower et al.(2009)]{Bower09} Bower, G. C., et al. 2009, \apj, 701, 1922. 
\bibitem[Bower et al.(2011)]{Bower11} Bower, G. C., et al. Radio Interferometric Planet Search. II. Constraints on Sub-jupiter-mass Companions to GJ~896A.  \apj, 740, 32 (2011)
\bibitem[Cant\'o et al.(2009)]{Canto09} Cant\'o, J., Curiel, S, \& Mart{\'\i}nez-G\'omez, E. A simple algorithm for optimization and model fitting: AGA (asexual genetic algorithm). \aap, 501, 1259 (2009)

\bibitem[Chabrier et al.(2023)]{Chabrier23} Chabrier, G., Baraffe, I., Phillips, M. \& Debras, F. 2023,  \aap, 671, A119.

\bibitem[Charbonneau et al.(2009)]{Charbonneau09} Charbonneau, D., et al. 2009, Nature, 462, 891. 
\bibitem[Curiel et al.(2011)]{Curiel11} Curiel, S., Cant\'o, J., Georgiev, L., Ch\'avez, C. E., \& Poveda, A. A fourth planet orbiting $\upsilon$ Andromedae. \aap, 525, 78 (2011)
\bibitem[Curiel et al.(2019)]{Curiel19} Curiel, S., Ortiz-Le\'on, G. N., Mioduszewski, A.J., \& Torres, R. M. Substellar Companions of the Young Weak-line TTauri Star DoAr21.  \apj, 884, 13 (2019)
\bibitem[Curiel et al.(2020)]{Curiel20} Curiel, S., Ortiz-Le\'on, G. N., Mioduszewski, A.J., \& Torres, R. M. An Astrometric Planetary Companion Candidate to the M9 Dwarf TVLM 513-46546.  \aj, 160, 97 (2020)
\bibitem[Curiel et al.(2022)]{Curiel22} Curiel, S., Ortiz-Le\'on, G. N., Mioduszewski, A.J., \& Sanchez-Bermudes, J. 3D  M Dwarf binary GJ~896AB 513-46546.  \aj, submitted (2022)
\bibitem[Dressing et al.(2015)]{Dressing15} Dressing, C. D., et al. 2015, \apj, 807, 45. 
\bibitem[Deshpande et al.(2012)]{Deshpande12} Deshpande, , R., Mart{\'\i}n, E. L., Montgomery, M. M.,, et al. 2012, \apj, 144, 99. 
\bibitem[Dupuy et al.(2010)]{Dupuy10} Dupuy, T. J., et al. 2010, \apj, 805, 56. 
\bibitem[Dupuy et al.(2015)]{Dupuy15} Dupuy, T. J., et al. 2015, \apj, 805, 56. 
\bibitem[Dupuy et al.(2016)]{Dupuy16} Dupuy, T. J., et al. 2016, \apj, 827, 23. 
\bibitem[Dupuy \& Liu(2017)]{Dupuy17} Dupuy, T. J., \& Liu, M. C. 2017, \apjs, 231, 15. 
\bibitem[Fernandes et al.(2019)]{Fernandes19} Fernandes, C. S., et al. 2019, \apj, 879, 94. 
\bibitem[Foreman-Mackey et al.(2013)]{Foremanmackey13}  Foreman-Mackey, D., Hogg, D. W., Lang, D., \& Goodman, J., emcee: The MCMC Hammer, \pasp, 125, 306, (2013)
\bibitem[Foreman-Mackey(2016)]{Foremanmackey16}  Foreman-Mackey, D., corner.py: Scatterplot matrices in Python, Journal of Open Source Software, 1, 24 (2016)
\bibitem[Forveille et al.(2005)]{Forveille05} Forveille, T., Beuzit, J.-L, Delorme, P. 2015, \aap, 435, L5. 
\bibitem[Forbrich \& Berger(2009)]{Forbrich09} Forbrich, J., \& Berger, E. 2009, \apjl, 706, L205. 
\bibitem[Forbrich et al.(2013)]{Forbrich13} Forbrich, J., et al. 2013, \apj, 777, 70. 
\bibitem[Forbrich et al.(2016)]{Forbrich16} Forbrich, J., et al. 2016, \apj, 827, 22. 
\bibitem[Gawronski et al.(2017)]{Gawronski17} Gawronski, M. G., et al. 2017, \mnras, 466, 4, 4211. 
\bibitem[Gillon et al.(2017)]{Gillon17} Gillon, M. et al. 2017, Nature, 542, 7642, 456. 
\bibitem[Green(1993)]{Green93} Green, R. M. (ed.) 1993, Spherical Astronomy (Cambridge: Cambridge Univ. Press)
\bibitem[Greisen(2003)]{Greisen03} Greisen, E., W.  In Information Handling in Astronomy: Historical Vistas, Vol. 285, ed. A. Heck (New York: Springer), 109 (2003)
\bibitem[Gizis et al.(2000)]{Gizis00} Gizis, J. E., Monet, D. G., Reid, I. N., et al. 2000, \aj, 120, 1085
\bibitem[Greisen(2003)]{Greisen03} Greisen, E., W.  In Information Handling in Astronomy: Historical Vistas, Vol. 285, ed. A. Heck (New York: Springer), 109 (2003)
\bibitem[Hallinan et al.(2008)]{Hallinan08} Hallinan, G., et al. 2008, \apj, 684, 644. 
\bibitem[Harding et al.(2013)]{Harding13} Harding, L. K., Hallinan, G., Boyle, R. P., et al. 2013, \apj, 779, 101
\bibitem[Hughes et al.(2021)]{Hughes21} Hughes, A. G., et al. 2021, \apj, submitted. 
\bibitem[Hunter(2007)]{Hunter07}  Hunter, J. D. Computing in Science and Engeneering. 9, 90 (2007)
\bibitem[Kalas et al.(2008)]{Kalas08} Kalas, P., et al. 2008, Sci, 322, 1345. 
\bibitem[Kennedy \& Kenyon(2008)]{Kennedy09} Kennedy, G.M., \& Kenyon, S.J. 2008, \apjl, 673, 502. 
\bibitem[Kennedy et al.(1997)]{Kirkpatrick97} Kirkpatrick, J. D., et al. 1997, \aj, 113, 1421. 
\bibitem[Kirkpatrick et al.(1995)]{Kirkpatrick97} Kirkpatrick, J. D., Henry, T. J., \& Irwin, M. J. 1997, \aj, 113, 1421
\bibitem[Konopacky et al.(2010)]{Konopacky10} Konopacky, Q. M., Ghez, A. M., Barman, T. S., et al. 2010, \apj, 711, 1087–1122.
\bibitem[Konopacky et al.(2012)]{Konopacky12} Konopacky, Q. M., Ghez, A. M., Fabrycky, D. C., et al. 2012, \apj, 750, 79
\bibitem[Kubas et al.(2012)]{Kubas12} Kubas, D., et al. 2012, \aap, 540, A78. 
\bibitem[Laughlin et al.(2004)]{Laughlin04} Laughlin, G. et al. 2004, \apjl, 612, L73. 
\bibitem[Newville et al.(2020)]{Newville20}  Newville, M., Otten, R., Nelson, A., et al. (2020). {lmfit/lmfit-py 1.0.1, 1.0.1 Zenodo}. {DOI 10.5281/zenodo.598352.}
\bibitem[Mayor \& Queloz(1995)]{Mayor95} Mayor, M., \& Queloz, D. 1995, Nature, 378, 355. 
\bibitem[McLean et al.(2012)]{McLean12} McLean, M., et al. 2012, \apj, 746, 23. 
\bibitem[Miles-Pa\'ez et al.(2017)]{Miles17} Miles-Pa\'ez, P. A., Pall\`e, E. \& Zapatero Osorio, M. R., 2017, \mnras, 472, 2297. 
\bibitem[Morales et al.(2019)]{Morales19} Morales, J. C., et al. 2019, Sci, 365, 1441. 
\bibitem[Muirhead et al.(2012)]{Muirhead12} Muirhead, P. S., et al. 2012, \apj, 747, 144. 
\bibitem[Ortiz-Leon et al.(2017)]{Ortiz-Leon17} Ortiz-Leon, G. N., Loinard, L., Kounkel, M. A., et. al. 2017, \apj, 834, 141.
\bibitem[Osten et al.(2009)]{Osten09} Osten, R. A., Phan-Bao, N., Hawley, S. L., Reid, I. N., \&
Ojha, R. 2009, \apj, 700, 1750
\bibitem[Phan-Bao et al.(2007)]{Phan-Bao07} Phan-Bao, N., Osten, R. A., Lim, J., Mart{\'\i}n, E. L., \& Ho, P. T. P. 2007, \apj, 658, 553
\bibitem[Phillips et al.(2020)]{Phillips20} Phillips, M. W., et al. 2020, \aap, 637, A38 (2020)
\bibitem[Pradel et al.(2006)]{Pradel06} Pradel, N. \& Lestrade, J. -F. Astrometric accuracy of phase-referenced observations with the VLBA and EVN.  \aap, 452, 1099 (2006)
\bibitem[Reid et al.(2014)]{Reid14} Reid, M.~J. \& Honma, M. Microarcsecond Radio Astrometry. \araa, 52, 339 (2014)
\bibitem[Reiners \& Basri (2009)]{Reiners09} Reiners, A. \& Basiri, G. \apj, 705, 1416 (2009)
\bibitem[Sahlmann et al.(2013)]{Sahlmann13} Sahlmann, J., et al. 2013, \aap, 556, A133. 
\bibitem[Sahlmann et al.(2016)]{Sahlmann16} Sahlmann, J., et al. 2016, \aap, 595, A77. 
\bibitem[Thompson et al.(2017)]{Thompson17} Thompson, A. Richard,  Moran, James M., \& Swenson, George W., Jr. Interferometry and Synthesis in Radio Astronomy, 3rd Edition (2017)
\bibitem[Torres et al.(2007)]{Torres07} Torres, R., Loinard, L., Mioduszewski, A. J., 
et al. 2007, \apj, 671, 1813 (2007)
\bibitem[van der Walt et al.(2011)]{Vanderwalt11} van der Walt, S, Colbert, S. C.. \& Varoquaux, G. Computing in Science and Engineering, 13, 22 (2011)
\bibitem[Wolszczan \& Frail(1992)]{Wolszczan92} Wolszczan, A., \& Frail, D. A. 1992, Nature, 355, 145. 
\bibitem[Zhang et al.(2020)]{Zhang20} Zhang, Q. et al. 2020, \apj, 897, 11.








\end{thebibliography}
\end{document}